\documentstyle[12pt,psfig,epsfig]{article}
\def\aprle{\buildrel < \over {_{\sim}}} 
\def\aprge{\buildrel > \over {_{\sim}}}   
\def\nubar{\overline {\nu} }
  
\begin{document}     

\title{Atmospheric neutrinos, long--baseline neutrino beams 
and the precise measurement  of the neutrino oscillation parameters}
\author{Giuseppe  Battistoni$^1$ and Paolo Lipari$^2$} 
\date{}

\maketitle

\begin{center}

{\small {\it 1. I.N.F.N., Sezione di Milano, via Celoria 16, I-20133 Milano, Italy}} \\
{\small {\it 2. Dipartimento di Fisica, Universit\`a di Roma ``la Sapienza'',  and}} \\
{\small {\it    I.N.F.N., Sezione di Roma, Piazzale A. Moro 2, I-00185 Roma, Italy}}

\end{center}

\begin{abstract}
Measurements of atmospheric  neutrinos  by Super--Kamiokande
an other detectors   have  given  evidence 
for the existence of neutrino oscillations
with large  mixing and $\Delta m^2$ in the  range
$10^{-3}$--$10^{-2}$~eV$^2$.  
In this  work  we  discuss critically  some  of  the possible
experimental strategies  to  confirm  this result and  determine
more accurately the  neutrino oscillation parameters.
A possible  method  is the 
development  of long--baseline accelerator neutrino  beams.
The accelerator beams can have higher intensity  and   
higher  average energy  than  the atmospheric  flux,
and   if $\nu_\mu \leftrightarrow \nu_\tau$ oscillations
are indeed the  cause of  the atmospheric  neutrino anomaly,  they can 
produce a measurable  rate of  $\tau$ leptons  for most 
(but not all) of the 
values of  the oscillation parameters  that   are a solution
to the atmospheric data.
On the other  hand 
measurements  of  atmospheric  neutrinos 
with large  statistics  and/or better experimental resolutions,
can  also provide   convincing evidence for  oscillations,
thanks  to unambiguous  detectable  effects  on the energy,
zenith angle and $L/E_\nu$ distributions  of the events.
The study  of these  effects  can  provide
a precise  determination of the oscillations  parameters.
The range of    $L/E_\nu$ 
available for atmospheric  neutrinos is  much larger than
in long--baseline accelerator experiments, and  the  sensitivity
extends  to  lower values of $\Delta m^2$.
\end{abstract} 

\vskip 3cm
\begin{center}
{\it To appear in the proceedings of the 1998 ``Vulcano
Workshop on Frontier objects in Astrophysics and
Particle Physics''.}
\end{center}
 
\newpage

\section{Introduction}

The data of  Super--Kamiokande \cite {SK}
has  strengthened  the evidence  for the existence of  an anomaly
in the flavor ratio of  atmospheric  neutrinos.
The indication for the  existence  of neutrino oscillation
has  originally appeared \cite{atm-problem}  as the measurement of a
ratio  $R_{\mu/e}$  of  $\mu$--like and $e$--like events
lower  that the  MC expectation.  The high  statistics data of
SuperKamiokande  show  distortions 
of the  angular  distributions  of sub--GeV and multi--GeV  $\mu$--like
that strongly support the neutrino oscillation hypothesis.
The angular  distribution of the $e$--like  events 
is  consistent with the no--oscillation  hypothesis,
in agreement  with the results  with  the  reactor experiment
Chooz \cite{Chooz}   that  essentially  rules  out the
$\nu_e \leftrightarrow \nu_x$ oscillations   in the interesting region
of parameter space.
It is  clear that   it is necessary to confirm
in an unambiguous  way if neutrino oscillations  are indeed
the cause  of the effects  seen in the atmospheric  neutrino
experiments. If this will turn out to be the case,
the next  task  is to measure with great accuracy the 
oscillation  parameters.

At present, the scientific community is discussing about the
best experimental strategy to accomplish this task.
Long--baseline (LBL) neutrino experiments are being designed for this
purpose, but there is also a strong debate on the role
of possible new atmospheric neutrino detectors different
from SuperKamiokande. The  main problem is  the fact
that the region  of  oscillation
parameters suggested by the  Super--Kamiokande data
($0.5 \le |\Delta m^2|/10^{-3}~$~eV$^2$~$\le 6$ at 90\% confidence level)
is lower than  previous  estimates,
and the planned LBL experiments could fail 
in achieving their goal of the 
unambiguous detection 
or  the ruling out of flavor oscillations  as  explanation
of the atmospheric  neutrino problem.

The purpose of this this work is to discuss and compare
the potential in proving unambiguously   the existence of oscillations
and in the  determination of the oscillation parameters, of both
long--baseline  (LBL) neutrino beams and
of  higher statistics and/or higher quality 
measurements of  atmospheric  neutrinos.
For  the sake of  clarity and  simplicity we will limit our discussion to 
the simplest solution for the problem, namely
the  existence of $\nu_\mu \leftrightarrow \nu_\tau$ 
oscillations. 
A general analysis in terms of a  3--neutrino scheme,
or considering also the possibility of  mixing with 
light sterile states  is of course desirable,
and should be considered in a more complete discussion.

The $\nu_\mu \leftrightarrow \nu_\tau$ oscillation probability   for  neutrinos
propagating in ordinary matter
is given    by the  well known  formula:
\begin{equation}
P_{\nu_{\mu} \to \nu_{\tau}}  = 
\sin^2 2 \theta ~\sin^2 \left ( {\Delta m^2 \over 4} \, {L\over E_\nu} \right)
\label{osc-prob}
\end{equation}
For small  values of  $L/E_\nu$, 
($L/E_\nu \ll |\Delta m^2|^{-1}$ )  expanding the second
sine in eq.(\ref{osc-prob})   the oscillation probability
can be approximated as:
\begin{equation}
P_{\nu_\mu \to \nu_\tau}  \simeq 
\sin^2 2 \theta \, (\Delta m^2)^2 \; {L^2 \over 16} 
\;{1\over E_\nu^2} 
\label{osc-prob-asym}
\end{equation}
note  that the two parameters
$\sin^2 2 \theta$  and  $\Delta m^2$ appear as a single
multiplicative  factor in the combination 
$\sin^2 2 \theta \, (\Delta m^2)^2$  and  cannot  be  measured  
separately.

This work is organized as follows:
in the next section we  discuss the  properties of the atmospheric neutrino
flux and of the proposed LBL  beams, including a short  analysis of 
systematic  uncertainties in the calculation of  no--oscillation predictions.
Then we discuss   the signatures
of neutrino oscillations  on   observable  quantities,
considering separately ``appearance'' and ``disappearance''  for LBL neutrinos.
We  then discuss the potential of atmospheric  neutrino measurements.
A summary and  discussion follow.

\section {Atmospheric and Accelerator Neutrino Beams} 

\begin{figure} [t]
\centerline {\psfig{figure=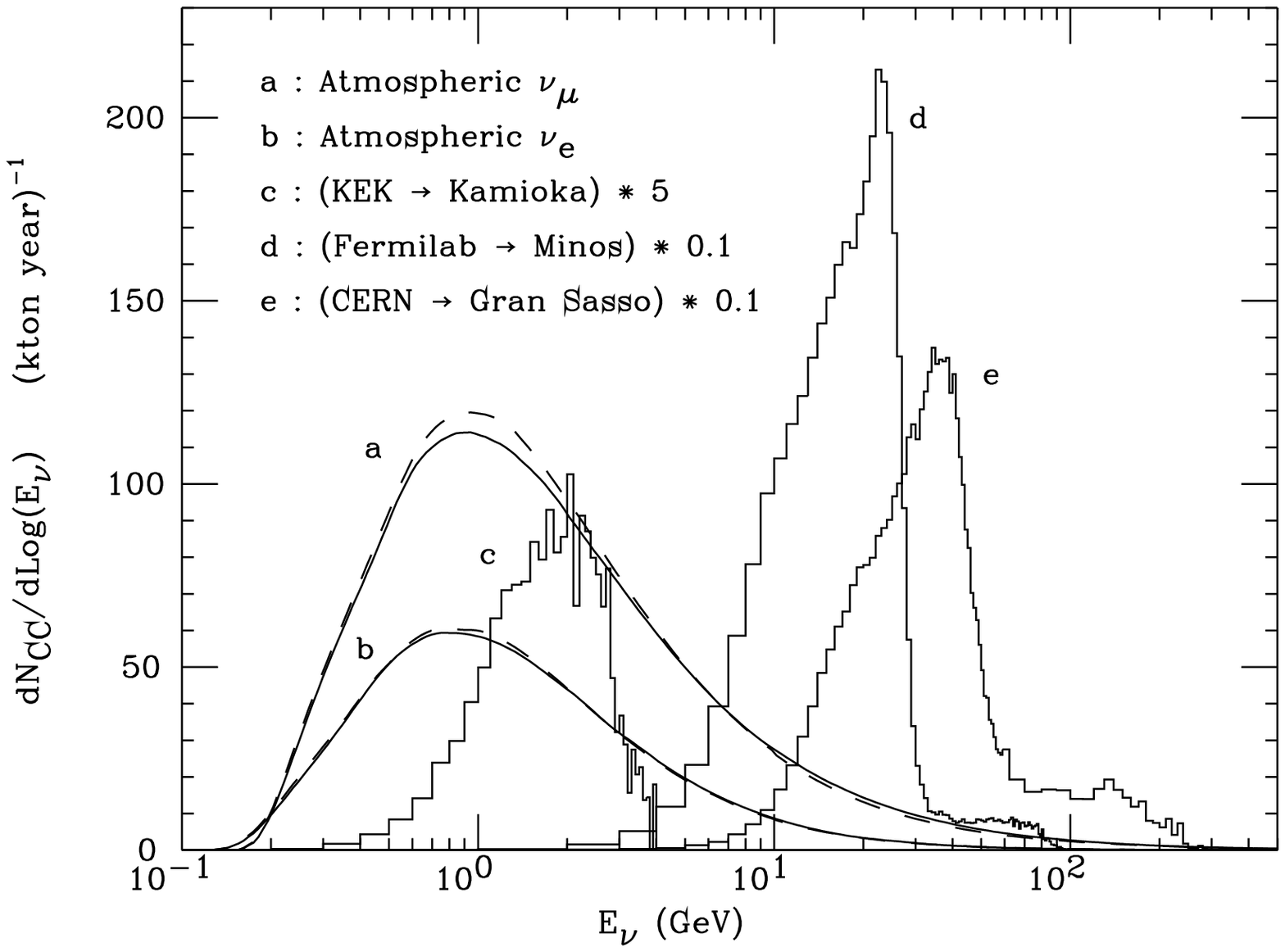,height=11cm}}
\vspace {0.25 cm}
\caption{Energy  distribution of  interacting (with charged  current)
atmospheric  neutrinos and antineutrinos,  and of  the  $\nu_\mu$ 
in three LBL  experiments.
All  calculations  assume the absence of  neutrino oscillations.
For atmospheric  neutrinos the solid (dashed) lines  are  calculated  with the
the Bartol  \protect\cite{Bartol} (Honda  et al. \protect\cite{Honda})
The scale of the vertical axis is absolute, note however   
that the LBL fluxes are  multiplied by  constant  factors.
\label{fig:spectrum} }
\end{figure}

 In fig.~\ref{fig:spectrum}
 we   compare  the energy  spectrum of atmospheric  neutrinos
with the predicted spectra   for  three  long--baseline  beams:
(i) KEK to SuperKamiokande   (K2K)  \cite{K2K}
(proton  energy $E_0 = 12$~GeV, neutrino path--length $L \simeq 250$~km,
intensity $N_p = 3.3\times 10^{19}$~yr$^{-1}$),
(ii)  Fermilab to MINOS   \cite{MINOS}
($E_0 = 120$~GeV, $L \simeq 730$~km,
$N_p = 2.0\times 10^{20}$~yr$^{-1}$)
and (iii) CERN to Gran Sasso laboratory  \cite{CERN}
($E_0 = 400$~GeV, $L \simeq 730$~km,
$N_p = 3.0\times 10^{19}$~yr$^{-1}$).
To estimate the proton intensities of 
the Fermilab and CERN beams  we  have used for both cases the  predicted 
repetition rate  of the accelerator and  considered a year   of 
220 days of running with an efficiency of 50\%.
The event  rates plotted in fig.~\ref{fig:spectrum} 
are calculated  in the absence  of  neutrino  oscillations, and
refer to charged  current (CC)
interactions of $\nu_\mu$'s    in the case of the accelerator  beams,
and $(\nu_\mu + \overline{\nu}_\mu$)'s  or 
$(\nu_e + \overline{\nu}_e$)'s  for atmospheric  neutrinos.
In all cases  we have  used   the same description of the neutrino  cross
section  from \cite{LLS}. The atmospheric  neutrino  event rate   
was calculated  using the Bartol  \cite{Bartol}
or Honda et al. \cite{Honda} neutrino  flux
for the  geomagnetic  location of  Kamiokande.
The event rates  are plotted 
on an  absolute scale (in (kton~yr)$^{-1}$), but to make the  plot 
more  readable  we have    rescaled  the KEK, Fermilab and  CERN beams
with  factors of 5, 0.1  and  0.1.  
Note that fig.~\ref{fig:spectrum} we  plot as  a  function
of the logarithms  of the energy the distribution
$dN_\nu^{cc}/d\log E_\nu$,  therefore the total rate is  proportional to
the area  under the curves or histograms\footnote{We  find that this   
representation  of the spectrum is  in most cases  more  useful  that the 
commonly used  plot of $dN_\nu/dE_\nu$  with a  linear  energy  scale.
For  example looking at  fig.~\ref{fig:spectrum} one  can easily read
the fraction  of the event rate   with $E_\nu  \le 10$~GeV 
(or  $E_\nu \ge 100$~GeV).}.

\begin{figure} [t]
\centerline {\psfig{figure=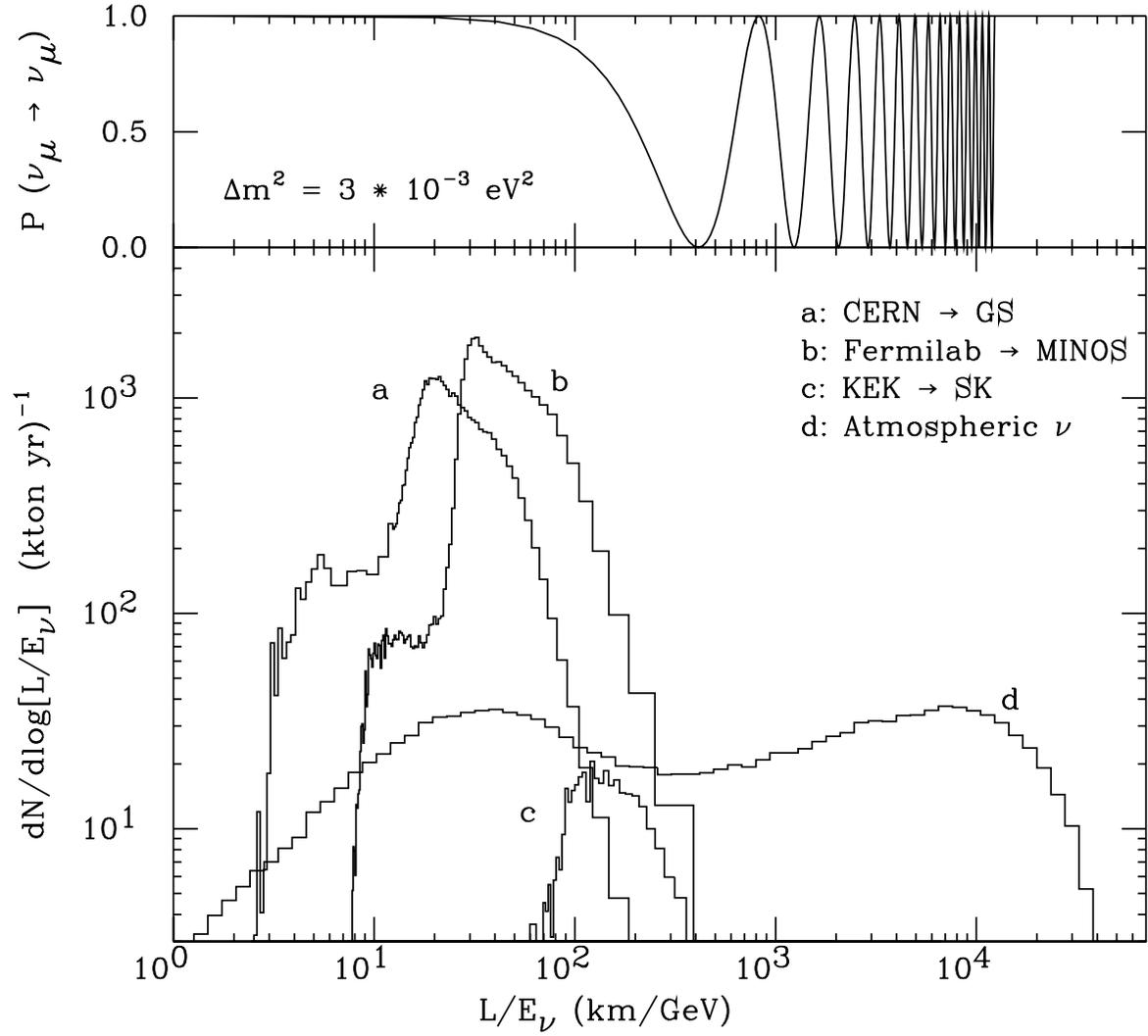,height=14cm}}
\vspace {0.25 cm}
\caption{Distribution in $L/E_\nu$  of the
charged current  events   expected  (in the  absence of  oscillations)  in 
three LBL experiments,  and for  atmospheric neutrinos 
with a cut $p_\mu \ge 0.2$~GeV.
In the upper panel  we  show the  oscillation probability 
for  maximal mixing and $\Delta m^2 = 3 \times 10^{-3}$~eV$^2$.
\label{fig:rate} }
\end{figure}

Neutrino  oscillations (in the absence  of  matter effects) are a
function  of the parameter  $L/E_\nu$  and it is therefore interesting to 
show the event  rate as a function of this   quantity.  This  is  done
in   fig,~\ref{fig:rate}   where we plot the  event rates  for 
atmospheric  neutrinos  and for the  3  LBL  experiments 
plotted as a function of the parameter
$L/E_\nu$.
The  event rates are calculated as in  fig.~\ref{fig:spectrum}
and are for   are  CC
interactions. In the case   of 
atmospheric  neutrinos the rate is for 
muon (anti)--neutrinos  and  we have  also included  the requirement that the 
final state $\mu^\pm$ has  a momentum $p_\mu > 200$~MeV.

In  the  upper panel of fig.~\ref{fig:rate}  we  
show an example  of the oscillation
probability (eq.\ref{osc-prob})
with values of the parameters 
$\sin^2 2 \theta = 1$  and $\Delta  m^2 = 3 \times 10^{-3}$~eV$^2$.
The oscillation parameters  of this example 
are chosen   as  `typical' values 
obtained  as solution of the neutrino atmospheric  anomaly, close  
to the point of minimum $\chi^2$  for the  combined  sub--GeV and multi--GeV
data of the SuperKamiokande detector \cite{SK}
($\Delta m^2 = 2.2 \times 10^{-3}$eV$^2$, $\sin^2 2 \theta=1$).
Previous  data of the Kamiokande  experiment \cite{Kamioka-atm}
suggested  a  higher  best fit value  
$\Delta m^2\simeq 1.5 \times 10^{-2}$eV$^2$.

Integrating over all   values of $L/E_\nu$ the event rates
for  atmospheric neutrinos,  and for the
K2K, Fermilab and CERN  experiments are
$N_\mu = 140$, 21, 2100  and 1530  (kton~yr)$^{-1}$.
The  two  proposed high energy 
experiment  obviously  benefit from  having a  large   event 
rate, 10  to 15 times higher than  atmospheric   neutrinos.

Atmospheric  neutrinos arrive from all  directions
including up--going  trajectories that have  crossed
the earth  have a very broad  range  of path--lengths 
$L \simeq 10$--10$^{4}$~km    that is  reflected  in a  broad
range  of $L/E \simeq 1$--10$^{5}$~km/GeV.  The 
plot of the   event rate as  a function of $L/E_\nu$  has
a characteristic  two--bumps  form, that  correspond
to down--going  and up--going particles. The `valley'  in  between is
populated mostly by particles  with directions
close to  the horizontal, the  event rate per unit
$L/E_\nu$  is lower in this region 
because the  path--length $L$ changes  rapidly  with the  
variation of the   neutrino  zenith angle $\theta_z$.
For accelerator neutrinos the   path--length $L$ is  fixed  
(the  width of the region where neutrinos are produced
$\Delta L \aprle 1$~km  being much smaller  than $L$)
and  the  distribution in $L/E_\nu$  reflects    the energy spectrum of the 
neutrinos (using a logarithmic  scale  
the   $L/E_\nu$  distribution is simply the energy distribution 
translated  and reflected).
The two  high energy  accelerator experiments have the same 
$L$ and the difference between the  two  distributions  reflects the
lower energy and higher  intensity of the  Fermilab    accelerator.
The  K2K experiment has  a  much lower intensity,  
was  carefully designed  to send  a controlled
flux of accelerator--made 
neutrinos  in precisely  the   range of values of
$L/E_\nu$ where neutrino oscillations  with parameters  suggested 
by the Kamiokande experiment  would have the greatest  visible
effects. 

Comparing the lower and upper panel of fig.~\ref{fig:rate}  
it can be noticed  the high energy  LBL beams  of Fermilab and  Minos
will have  a limited sensitivity   to  oscillations with
$\Delta m^2 \aprle 3\times 10^{-3}$~eV$^2$   because of the small
expected  event rate  $L/E_\nu \aprge 100$~km/GeV.
The  probability $P_{\nu_\mu \to \nu_\mu}$ has  
minima  for values  $L/E_\nu = n\;(L/E_\nu)^*$   where $n$ is an integer.
The first minimum  occurs at 
\begin{equation} 
\left ({L \over E }\right )^*
 = {2 \pi \over \Delta m^2} = { 1236  \over \Delta m^2_{-3} }
~{ {\rm km/GeV} }
\end{equation}
where $\Delta m^2_{-3}$ is the value of the squared mass difference
in units of $10^{-3}$~eV$^2$.
It is  clearly very desirable to  measure  the neutrino oscillation probability
up  to values of $L/E_\nu$    that approach and exceed  $(L/E)^*$.
If this is  not achieved,
the value of  the mixing parameter $\sin^2 2 \theta$   that  gives  the
depth of the  minimum in the survival probability  cannot  be 
measured,  and in fact the phenomenon   of neutrino {\em oscillations}
cannot be experimentally proven.

In LBL  experiments the path--length $L$ is  fixed  and 
the $P_{\nu_\mu \to \nu_\mu}$ probability has  minima 
for   $E_\nu = E^*/n$   with 
$E^* = \Delta m^2  L /(2 \pi)$.  
For the  proposed  LBL  experiments the highest energy minimum is:
\begin{equation}
E^* ({\rm K2K }) = 0.20 ~\Delta m^2_{-3} ~{\rm GeV}
\end{equation}
\begin{equation}
E^* ({\rm CERN}) = 0.59  ~\Delta m^2_{-3} ~{\rm GeV}
\end{equation}
For  the Fermilab experiment the  no--oscillation CC event  rates
with $E_\nu$ below 10, 5 and 2 GeV are  420, 50 and 2.5  events per year. 
For the present design of the CERN   beam
with lower  intensity  and  the focus  optimized for  higher  energies,
the event rates  are roughly 34, 3.5 and 0.5 ~yr$^{-1}$.
Given  these rates the $L/E_\nu$  region  where one expects the 
survival  (transition) probability   to have a minimum (maximum)
appears to be  accessible only   for the high values  of $\Delta m^2$  
in the range suggested  by the  atmospheric  neutrino data.

It is  possible   to prove the existence of
flavor  transitions   without  having neutrinos with $L/E_\nu$ as  large
as  $(L/E)^*$.    In this  case  only the product
$\sin^2 2 \theta \, (\Delta m^2)^2$ is  
experimentally  measurable.
For $\Delta m^2 \aprle 10^{-2}$~eV$^2$ this is what 
will happen  with the  proposed  LBL neutrino  beams.

\subsection {Atmospheric neutrinos}

Detailed calculations of atmospheric neutrino fluxes 
can be found in literature. At present, two of the most
quoted results are those of 
Honda et al. \cite{Honda} and those of the Bartol group \cite {Bartol}.
These neutrino flux   calculations 
are  based on a unidimensional  description of the c.r.  showers.
Recent  comparisons  with a 3--dimensional  calculation based on
the FLUKA  montecarlo \cite {FLUKA} confirm  that the approximation
is adequate at least for  neutrinos above 200~MeV.

We refer to the quoted references for the description of the main features
of atmospheric neutrinos. 
Here we limit ourselves to focus some important properties which
are relevant in the search for oscillations and that can be predicted
very reliably.

\begin {enumerate}
\item The neutrino fluxes are   to a very good approximation
up--down symmetric:
\begin {equation}
\phi_{\nu_{\alpha}} (E, \cos \theta_z ) = 
\phi_{\nu_{\alpha}} (E, -\cos \theta_z )
\label{zenith-symmetry}
\end{equation}
\item The electron neutrino and muon neutrino fluxes
originate from the decay  of the same   charged
mesons,  and are strictly related  to each other:
\begin {equation}
\phi_{\nu_{e} + \nubar_{e}} (E, \cos \theta_z ) =  r_{e \mu} 
~\phi_{\nu_{\mu} + \nubar_\mu } (E, \cos \theta_z )
\end{equation}
where the factor  $r_{e \mu}$ is only  weakly dependent  on  the neutrino
energy and direction  and close to ${1 \over 2}$.
\end{enumerate}
Geomagnetic  effects   introduce at low  energy a small
violation of the   up--down symmetry. 
of order of $\sim 10\%$   for the  `sub--GeV'  sample of SuperKamiokande,
but  in the absence
of  oscillations  the  relation
(\ref{zenith-symmetry}) can be reliably predicted 
as valid  with  few percent   accuracy above  
 1~GeV \cite{geomagnetic-effects}. 
The size of the geomagnetic effects  on neutrinos can   also 
be measured   observing a small east--west   effect  
\cite{geomagnetic-effects} that is  approximately
independent from oscillations.
No mechanism   besides neutrino oscillations, has  been  
proposed  to  explain the large  deviation   observed 
from the up--down symmetry observed in 
the multi--GeV  data  of SuperKamiokande.

The deviation of the   measured 
$e$--like/$\mu$--like  ratio from the expected value  has been the
origin of the  atmospheric  neutrino  problem.
In the presence of $\nu_\mu \to \nu_x$ oscillations  
with $x \ne e$ the electron neutrino  flux   can  be a 
quite accurate monitor  of the no--oscillation flux.

\subsection {Long baseline accelerator neutrino beams}
It can be useful to have a qualitative  understanding of the 
intensity  and energy  distribution of a long baseline accelerator beam.
To  a  good  approximation  (a more  complete  discussion   is  contained
in the appendix) the    energy distribution of the 
flux  of neutrinos  at the  far  detector  is given by the following
formula:
\begin{eqnarray}
{dN_{\nu} \over  dE_\nu } (E_\nu) \simeq 
{N_p \over 4 \pi \, L^2} & &  \left \{
2.34 \;
~\left[ 
{dn_\pi \over dE_\pi} (E_\pi,E_0) 
 ~P^{\pi}_{dec} (E_\pi)\right]_{E_\pi = 2.34\,E_\nu} 
~{E_\nu^2 \over \langle p_{\perp\nu}^\pi (E_\nu)\rangle^2 }
\right. \nonumber \\
 & & \left . 
  +  1.05\;
~\left[ 
{dn_K \over dE_K} (E_K,E_0) 
 ~P^{K}_{dec} (E_K)\right]_{E_K = 1.05\,E_\nu} 
~{E_\nu^2 \over \langle p_{\perp\nu}^K (E_\nu)\rangle^2 }
\right \}
\label{eq:flux-LBL}
\end{eqnarray}
where
$E_0$ and $N_p$ 
are the  energy and intensity (particles  per unit time)
of the primary proton beam,   $L$  is the distance between 
accelerator  and  detector,
$dn_\pi/dE_\pi$  
($dn_\pi/dE_K$) is the inclusive   energy  distribution of
positive  pions  (kaons) in  the  final state, and
$P_{dec}^{\pi(K)}(E_\pi)$ is the   decay  probability  
in the $\mu \nu_\mu$   final state  of  charged  pions
and  kaons that  depends   
on the  dimensions  of the decay  volume  after the interaction  region
(in  the  case of the kaons   one  has 
also  to include the branching  fraction
$B_{K\to \mu\nu} \simeq 0.64$).
The   numerical   factors $2.34 = m_\pi^2/(m_\pi^2 - m_\mu^2)$ and
$1.05 = m_K^2/(m_K^2 - m_\mu^2)$  are due to the  fact that  the neutrinos
produced in the direction   parallel to   parent  particle  
momentum in the $ \mu\nu$  decay   of a relativistic  pion (kaon)
have momentum $E_\nu = E_\pi/2.34$ ($E_\nu  = E_K/1.05$).
Because of this  kinematical  reason,
the  neutrinos  produced in pion (kaon) decay have a spectrum  that  extends
up to different maximum  energy.
The factor  $E_\nu^2/\langle p_{\perp\nu}\rangle^2$  is  due to the fact
that the intensity of the neutrino  flux  is  more intense  when the neutrinos
are  emitted in a small solid  angle.
Note that the  transverse  momentum of the  neutrinos
has  a contribution due to the $p_\perp$ obtained
in the   decay of the parent  particle, and a   contribution due to
the  $p_\perp$ of the parent  particle:
\begin {equation}
p_{\perp\nu} = p_\perp^{\pi \to \nu} \oplus p_\perp^\pi
\end{equation}
The  second  contribution can  be  reduced  optimizing the   focusing 
downstream  of the target,  while the   first one  is  unavoidable.
Since  the maximum  transverse   momentum   in a pion (kaon) decay is
29.8  (236) MeV, the contribution of the kaons is   suppressed  with respect to
the  pion one.

In the approximation of  `perfect--focus'  it is assumed  that  all   charged
mesons  of the appropriate  charge  and  positive  longitudinal  
momentum, after the  focusing  system  have  the 3--momentum parallel
to  the primary proton   beam  (that is 
$p_\perp^{\pi,K} \equiv 0$), neglecting  also    secondary interactions
and  decays  before  or  during the magnetic  bending.
In this  case   (see eq.\ref{eq:perfect-focus0})  it is possible to
deduce the  validity of  formula (\ref{eq:flux-LBL}) rigorously.
More in general  all  details of  the  focusing  system  are  absorbed 
in the  definition of $\langle p_\perp^{\pi,K} (E)\rangle$.

The number of  pions of energy $E_\pi$
produced in an interaction  by a primary proton of  energy $E_0$
is well  represented  by the scaling  expression:
\begin{equation}
 {dn_\pi \over dE_\pi} (E_\pi, ~E_0)  \simeq {1 \over E_\pi }
~F \left ( {E_\pi \over E_0} \right )  
\label{eq:scaling}
\end{equation}
where the function $F(x)$ is approximately independent
from the primary energy  $E_0$,  and decreases monotonically
from a finite value for $x \to 0$,   to zero for $x\to 1$\footnote{Very roughly
$F(x) \sim  (1-x)^4$.}.
Therefore the number  of pions  at {\em all}  energies increases with
increasing $E_0$.      
The known and observed violations of the  scaling expressed
by Equation (\ref{eq:scaling}),
result  in a higher  rapidity  plateau  and in   more rapidly increase of
the multiplicity  of  low  energy pions and kaons.

We can  make some  general  considerations  on the flux
\begin {enumerate}
\item  The neutrino flux  
grows  linearly   with the  intensity of the 
proton beam, and decreases  with  distance  as $L^{-2}$.

\item  In the perfect focus  approximation 
at low  energy the spectrum of the neutrinos
as approximately the form  $\phi_\nu \propto E_\nu$, 
and the   energy distribution of  
the interacting  neutrinos
is  $dN_{int}/dE_\nu \propto E_\nu^2$.

\item Increasing the  energy $E_0$ of the proton beam, for the same
target and   focusing  configuration, the neutrino flux increases for
all  neutrino  energies $E_\nu$  (and not  only  for the  highest energies).
However  one    should  also  consider the fact that the  intensity
of the proton  beam can increase to the shorter time
needed  to  accelerate  protons at a lower  energy.
\end{enumerate}

\begin{figure} [t]
\centerline {\psfig{figure=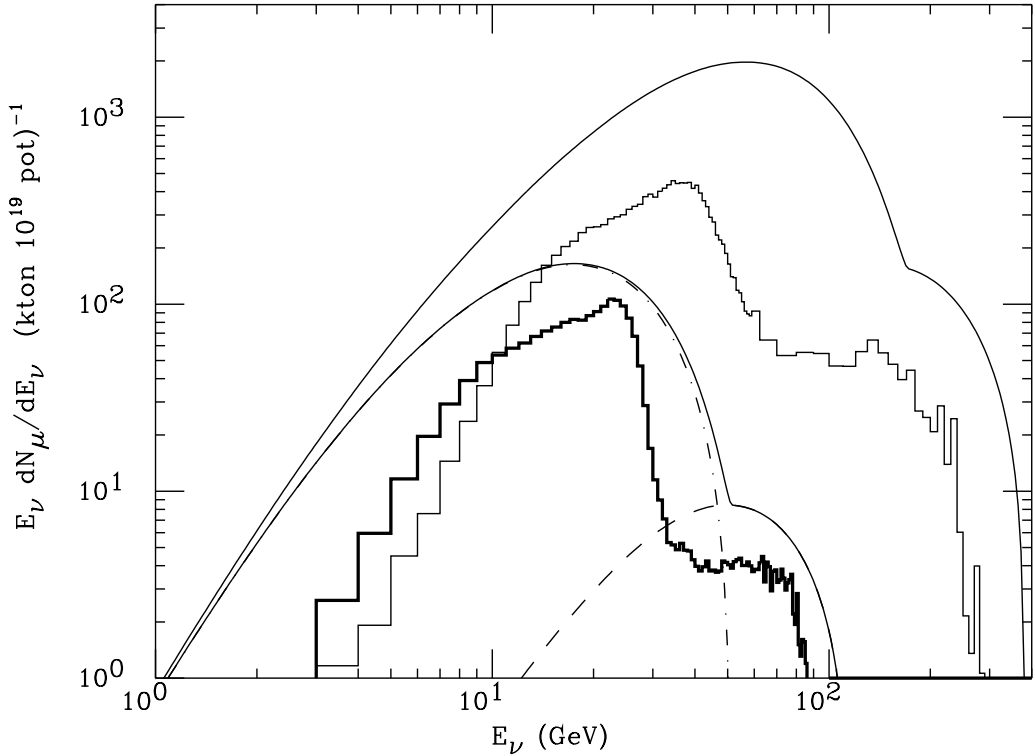,height=10cm}}
\vspace {0.4 cm}
\caption{ Energy distribution  of the   interacting  neutrinos
(charged current interactions)  for the Fermilab to
Minos  (thick histogram) and 
CERN to Gran Sasso (thin histogram)  LBL  beams,  compared  with an
analytic  approximation under the assumption of
perfect  focusing. 
The energy of the proton  beam  is  120  and  400 GeV  for the two cases.
The absolute scale of the  rate is  per kiloton and for  a  fixed  number
($10^{19}$) accelerated protons on target.
\label{fig:ideal} }
\end{figure}

In fig.~\ref{fig:ideal}  we   show  a calculation
of the rate of CC interactions    
$\phi_{\nu_\mu} (E_\nu;~E_0,L) \, \sigma_{CC}^{\nu_\mu}$ 
calculated  under the approximation of perfect  focusing
(equation (\ref{eq:perfect-focus0}))  for
$L= 730$~km and  $E_0 = 120$ and 400~GeV and  compare with the
results  of the   complete  Montecarlo simulations  of the 
Fermilab and CERN LBL beams.
For the decay probability we have  considered 
a decay tunnel  of length $T$ and a large  transverse section.
Then $P_{dec}(E_a) = 1 - \exp [-\varepsilon_a/(\beta \,E_a)]$,
with   $\varepsilon_a = m_a~T/\tau_a$. For $T = 800$~m
$\varepsilon_{\pi(K)}  = 14.3$~(106.5)~GeV.
For the  lowest energy (120~GeV)  we  also  show 
the  contribution of $\pi^+$   (dot--dashed)  and $K^+$ 
(dashed line).
The kaon   contribution  extends  to higher energy   because
a   neutrino   emitted in the forward direction in a kaon decay
has  energy $E_\nu = 0.954~E_K$  
while in  a pion decay  the neutrinos   get at most a
fraction  $0.427$  of the  parent energy.
The kaon contribution is suppressed with respect to the pion's one
because of the  larger $p_\perp$  available in the decay,
but  enhanced at  large  energy   because of the shorter lifetime.

\subsubsection {Optimization of LBL neutrino  beams}

The  construction  of a LBL 
neutrino beam  involves   choices for $L$, the distance  between the 
accelerator  and the detector, $E_0$  the energy of the  primary  proton beam,
and the detailed  design of the target and   magnetic focusing  system.

Of course  there are  technical  and  `financial' constraints
on  the design,  for  example 
the   choice  of $L$ is   strongly  limited  by the location 
of the existing  particle accelerators  and the possible  sites  for 
the neutrino detector.   
The `optimum'   design  of the neutrino  beam  depends 
on the   channel of  oscillations  that is  searched  for  
(for  example $\nu_\mu \to \nu_\tau$  or $\nu_\mu \to \nu_e$), the 
the region of  parameter space that is   searched for
oscillations  (for example  large or small $|\Delta m^2|$),
and the  experimental  strategy  used  in the search
(for example  `appearance'  and  `disappearance'  experiments).

Therefore if  several  detectors  using  different techniques
or  having  different  aims  are  planning to use the same    neutrino beam,
it will be  necessary to  find  a  compromise solution between  different
optimum  designs.   Ideally  the   design   of the neutrino  beam
and  of the detectors  (and their  search  strategies) 
should  be  developed   at the same time.

A  quantity that is  often  quoted 
in the  discussion of   the design of  a  neutrino  beam is
 the rate  of  charged  current muon neutrino interactions 
expected  in the absence of oscillations ($N_\mu^\circ$):
\begin{equation}
N_\mu^\circ = 
\int dE_\nu \; \phi_{\nu_\mu}^\circ  (E_\nu) \, \sigma_{\nu_\mu} (E_\nu)
\end{equation}
It should  be stressed  that the 
optimization of  the neutrino    beam  in  general  does {\em  not}
coincide with the maximization of $N_\mu^\circ$.

For  example, 
in the case of experiments  that  are  searching  for the 
appearance of $\tau$ leptons,  a more interesting 
quantity  is the  rate  $N_\tau$ of  charged  current  interactions  
of $\nu_\tau$\footnote{For  a  realistic  discussion one  should also
include  a  discussion  of  an   energy dependent  detection efficiency,
and  also  the possible sources  of background.}.
The rate  $N_\tau$  depends  of  course on the oscillation  parameters:
\begin{equation}
N_\tau (\Delta m^2, \sin^2 2 \theta)  = 
\int dE_\nu \; \phi_{\nu_\mu}^\circ  (E_\nu) 
\; P_{\nu_\mu \to \nu_\tau} (E_\nu, \Delta m^2, \sin^2 2 \theta) 
\; \sigma_{\nu_\tau} (E_\nu)
\label {eq:rate-tau}
\end{equation}

It is interesting to discuss the   form  that 
this  equation   takes in the limits of 
very large  or very small $|\Delta m^2|$, or more precisely for 
$|\Delta m^2|~L/E_\nu$  much  greater   (smaller) than unity.

In the limit of  large $|\Delta m^2|$, averaging  over the rapid
oscillations the transition probability 
takes an  energy independent   constant value
$P_{\nu_\mu \to \nu_\tau}  \simeq {1\over 2} \, \sin^2 2 \theta$  and we have:
\begin{equation}
N_\tau \simeq {\sin^2 2 \theta \over 2} \; n^\infty_\tau 
\end{equation}
where 
\begin{equation}
n^\infty_\tau  = 
\int dE_\nu \; \phi_{\nu_\mu}^\circ  (E_\nu) 
\; \sigma_{\nu_\tau} (E_\nu) = 
{\langle \sigma_{\nu_\tau} \rangle \over 
\langle \sigma_{\nu_\mu} \rangle} 
\;  N_\mu^\circ 
\end{equation}
$\langle \sigma_{\nu_{\tau(\mu)}} \rangle$  is the   charged current $\nu_\tau$
($\nu_\mu$) cross section averaged over the neutrino  flux spectrum.

In the limit of  small $|\Delta m^2|$  
the oscillation  probability can be  approximated   with the form 
of eq.(\ref{osc-prob-asym})
and the rate  of $\tau$ production  becomes:
\begin{equation}
N_\tau \simeq 
\sin^2 2 \theta \, (\Delta m^2)^2 \; n_\tau^0 
\label{eq:rate-tau-0}
\end {equation}
with 
\begin{equation}
n^0_\tau = {L^2 \over 16} 
~\int dE_\nu  
~{ \phi_{\nu_\mu}^\circ (E_\nu) \over E_\nu^2} \; \sigma_{\nu_\tau} (E_\nu)
\label{eq:nutau0}
\end{equation}
There  are several points that  are  worth  discussing  about
equations (\ref{eq:rate-tau-0},\ref{eq:nutau0}) 
 and the limit of small $|\Delta m^2|$:
\begin {enumerate}
\item The    limit  of  small $|\Delta m^2|$ 
is  to a  good  approximation  valid  for the planned
LBL  neutrino   beams  and the range of $|\Delta m^2|$    indicated  by
Super--Kamiokande.

\item In this  limit the rate   of $\tau$ production  grows  
linearly with  $\sin^2 2 \theta$ and {\em quadratically}  with 
$|\Delta m^2|$.

\item  The  oscillation parameters  enter  in the  rate  as  a single 
energy dependent multiplicative  factor:
$R_\tau \propto \sin^2 2 \theta \, (\Delta m^2)^2$.  Therefore  the  
two parameters  cannot  be  disentangled from each other.

\item In the evaluation of  the quantity
$n^0_\tau$ 
(equation (\ref{eq:nutau0})) 
each neutrino  energy  is  weighted  by the 
intensity of the flux, the  cross section for $\tau$  production and
an  additional  factor  $E_\nu^{-2}$  that takes  into  account
the oscillation  probability that  decreases  with  increasing  energy.
The  most `useful'  neutrino energies  are  those not  much  higher than
the   $\tau$--lepton threshold. 
The optimum spectrum is significantly softer than one optimized 
spectrum  for  a search  for  $\tau$ appearance at large $|\Delta m^2|$.

\item The   measurement   of the combination 
$\sin^2 2 \theta \, (\Delta m^2)^2$   depends on the   measurement of the
{\em absolute} rate of $\tau$--production. Therefore uncertainties on
the   absolute value of the $\nu_\tau$ cross section,  the 
detection efficiency  and  the neutrino flux  intensity
are important  sources of  systematic  uncertainty.

\item The  rate $N_\tau$  is {\em independent} on  $L$.
The  neutrino flux  at the far detector 
decreases as  $L^{-2}$  because of the divergence of the beam,   but  the
oscillation probability grows proportionally to $L^2$.  The resulting rate 
is independent from $L$.
\end {enumerate}

For disappearance  experiments
one  has  to maximize the   rate difference 
$\Delta N_\mu = N_\mu^\circ - N_\mu$
between the  observed rate and the no--oscillation rate, or 
the ratio $\Delta R_\mu/R_\mu^\circ$. 
The  crucial  region is the one   to the energy
\begin{equation}
 E^* = {\Delta m^2 \, L \over 2 \pi} = 0.59 \; 
\left ( {L \over 730~{\rm km} } \right ) ~
\left ( {|\Delta m^2|  \over 10^{-3}~{\rm eV}^2 } \right ) ~{\rm GeV}
\end{equation}
Neutrinos  with  energy $E_\nu \gg E^*$ do not oscillate
and   therefore are useless in  the search  for  oscillations 
and  a potential source of  background.

Summarizing, the optimization 
of a neutrino beam depends  on the experiment  one wants to perform.
This poses some  difficult problems  if experiments
with different goals  or using different  strategies   want to  use the same
beam.
A neutrino beam optimized for a short--baseline  $\nu_\tau$ appearance  
experiment  such  as  COSMOS  or TOSCA   will have higher energy 
than   a  beam  optimized for the ICARUS or OPERA experiments 
searching for  oscillations  at low $|\Delta m^2|$. 
The sensitivity of  a disappearance  experiment   is 
optimized   constructing  a neutrino beam  of  still lower  energy.

\subsection {Uncertainties in the calculation of  the $\nu$ beams}

Comparing the LBL accelerator  beams   with the atmospheric  flux
we can  note that  conceptually  the  process   of neutrino production
is  the same for both   processes.  A primary  high  energy hadron  beam
interacts  with  a target,  and  neutrinos are produced in the  
weak decay  of final state  mesons.

The primary  flux is  for  all practical purposes exactly known  
in an accelerator experiment,  while   there are  significant 
uncertainties in the   normalization ($\sim 15$\%) and   energy dependence
of the primary  cosmic ray (c.r.) flux (see  \cite{flux-comparison}
for a detailed  discussion). At low   energy it is  also 
necessary to   include  a  treatment of the 
solar modulation and of the geomagnetic  effects.
Time-varying magnetic fields in  interplanetary space
cause a time  dependence of the c.r. primary flux at low energies.
The geomagnetic field prevents the low  rigidity particles from
arriving at the surface of the earth,   resulting in  
a primary flux that  is not  isotropic and is  different 
for different  locations.  Both effects  however
vanish at sufficiently high energy.

Uncertainties in the modeling of particle  production  in  hadronic  showers
affects the calculation  of the flux
for both accelerator  and atmospheric  neutrinos. 
For atmospheric neutrinos  one has  also to consider the
interaction of  nuclear projectiles   (that account   for
a fraction  of few percent of the neutrinos)  and of weakly decaying
mesons (a small  contribution to the flux),
and    describe the interactions
of primary particle with energy  in a broad interval 
$E_0 \sim 5$--50~GeV (for   neutrinos  with energy from 300 MeV to 
2 GeV \cite{Bartol}. However since both the primary flux and
the resulting neutrino flux are  approximately isotropic 
only the energy distribution of the final particles is  important.
In the case of LBL accelerator beams also the 
transverse  momentum distribution  and 
the correlation between transverse and longitudinal  momentum 
of the  secondary particles  is critically important.
Detailed experimental  studies  of  particle  production
in hadronic  interactions  of  the appropriate energy (the  
beam  energy for accelerator   experiments and 
$E_0 \sim 10$--30~GeV  for atmospheric  neutrinos)
could reduce the uncertainties
in the calculation of the neutrino  spectrum and  normalization.

The description of particle  transport in 
the target and decay volume is another  source of uncertainty
that is important  especially for accelerator  neutrinos.
In the case of a LBL accelerator beam  the structure of the 
target and of the  magnetic focusing system must be 
known  and described  in great detail. The detector subtends  a very small
solid  angle  and a  very accurate and  detailed 
calculation of the  angular divergence 
of the neutrino  beam  as a function of energy is essential.
The c.r. showers develop in a  medium with small and 
slowly  varying  density.  There are small uncertainties 
related  to the time variations  and geographical position dependence
of the  atmospheric  density that
in the current simulations  is simply 
described as the average $\rho(h)$.
The polarization of muons  produced in two  body decay and  
its effect  on the spectra of neutrinos  coming   from chain  decay
as $\pi \to \mu \to \nu$  is  taken into account in  the
atmospheric  neutrino calculations  but the possible  effect  of 
muon depolarization  before decay 
is usually neglected.
This  is  believed to be a 
good approximation; in any case  the main effect  of  the 
polarization  is  a suppression (enhancement)   of the 
muon (electron) neutrino  flux of $\sim 10\%$, and the  uncertainty  related to
this  effect  cannot  be  large.

Both for accelerator and atmospheric neutrinos there are
methods to control the calculation of the no--oscillation
neutrino fluxes.
In an accelerator  it is  of course  possible to  
construct a near detector.   This is  indeed  an   very  important
possibilities that is  essential  for  any disappearance  experiment.
The main difficulty of this  approach is that  
the solid  angle  subtended by the far  detector  corresponds
to  a region  with transverse size $\sim 1$~cm for  a near detector
at the distance of one kilometer, and  the event rate 
in this  region is small (by definition equal to the far detector rate).
Therefore in practice the 
monitoring of the beam  will require  some assumptions
about the  angular distribution of the neutrinos.
It is expected  that
beam  intensity at the far detector will  be  approximately constant 
in a region of radius 100--200  meters, much 
larger that the detector,
with a correspondingly higher tolerance in the pointing  accuracy.
Therefore it should be possible to  use  an  angular region
much  broader than the  far detector  solid angle   to monitor the beam in
a near detector. This assumption can be, at least in
part, tested experimentally.
In studies  of  detector  performance, the  
fact that the relevant events in the near detector are  all
close to the detector  axis, while in the far detector 
are   uniformly spread  in the   plane  transverse to the beam
is a potential bias  that  requires a correction.

In summary we would like to  note  that 
according to a rather widespread `common 
wisdom' the systematic uncertainties  on 
the flux of accelerator neutrinos are much smaller than
for atmospheric neutrinos. 
This statement  should be qualified because is not in general correct.
In an accelerator there is the possibility of 
a close  detector, and of 
sophisticated monitoring systems.
In atmospheric  neutrinos  one can use the up--down symmetry
and the   relation between the $\nu_\mu$ and $\nu_e$  fluxes
as  essentially model independent tools  to estimate the 
no--oscillation  rates.

\section {Appearance Experiments}

In some  sense the most  convincing  evidence for the  existence of neutrino
flavor oscillations  is  the positive  identification of the 
charged  current  interactions  of $\tau$ neutrinos in the detector.
The unambiguous  detection of such events would   be   unquestionable proof
of oscillations, since  direct production of $\nu_\tau$'s  is  
predicted  (and  indeed  measured by CHORUS  and NOMAD) to be negligibly small.

\begin{figure} [t]
\centerline {\psfig{figure=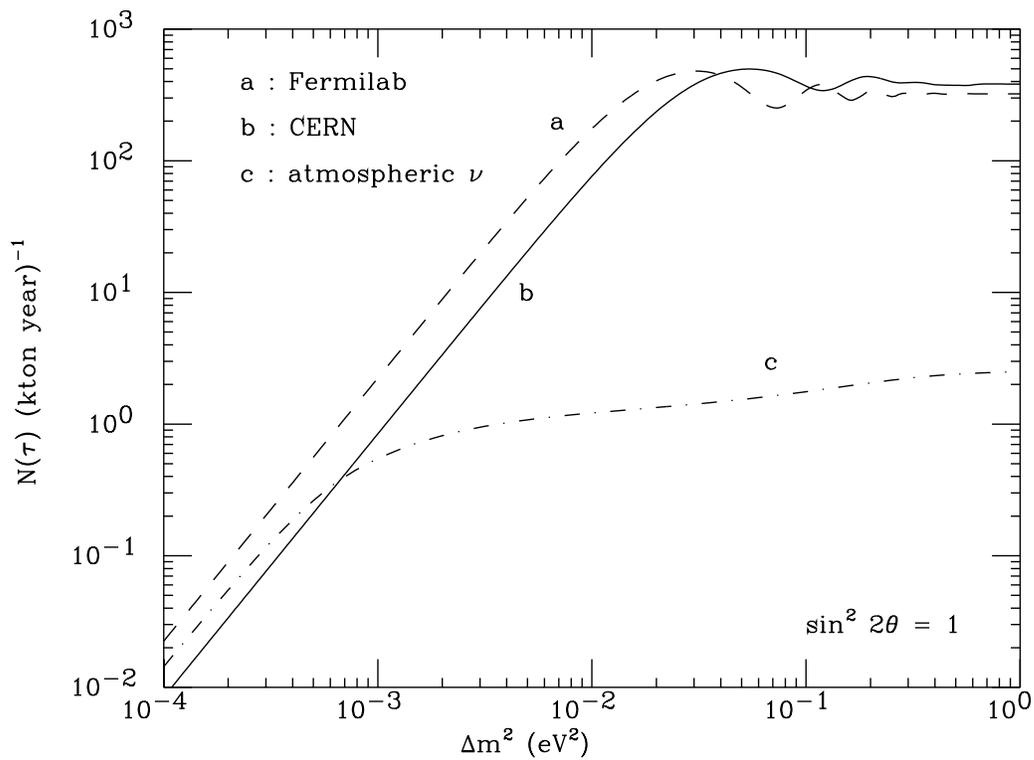,height=10cm}}
\vspace {0.4 cm}
\caption{ Rate of $\tau$ lepton  production as  a  function
of $\Delta m^2$  assuming  maximal  mixing
$\nu_\mu \leftrightarrow \nu_\tau$ oscillations.  \label{fig:tau} }
\end{figure}

In fig.~\ref{fig:tau} we show as a function of $\Delta m^2$ the event
  rate of charged current $\nu_\tau$ interactions calculated assuming
  the existence of $\nu_\mu \leftrightarrow \nu_\tau$ oscillations
  with maximal mixing.  The different curves are for atmospheric
  neutrinos and for the reference beams of the Fermilab and CERN long
  baseline projects.
  Essentially all  neutrinos  in  the K2K project  are 
  below   threshold for $\tau$ production  and the $\tau$--event rate
  is  vanishingly small in this case..

The rate of $\tau$  production   by  atmospheric  neutrinos
remains small, of order  1--2~events per (kiloton~yr)  even in the presence 
of large  mixing between muon and tau neutrinos. This is a  consequence
of the fact that
atmospheric  neutrinos  have  a  soft  spectrum and  
only a small fraction is  above the energy threshold
for $\tau$ production.
 
  The  high energy beams of Fermilab and CERN
can provide  a significant rate of $\tau$'s. 
Looking at fig.~\ref{fig:tau} one can  distinguish three
regions of $|\Delta m^2|$  where 
the rate of $\tau$  production  (for  a  constant  value  of the mixing
parameter) has a different behaviour:
\begin{enumerate}
\item for $|\Delta m^2| \aprle 10^{-2}$~eV$^2$   the rate grows as
$N_\tau \propto (\Delta m^2)^2$,
\item in an intermediate  region it oscillates
\item for large  $|\Delta m^2|$    becomes  a constant.
\end{enumerate}

For  the same number of protons on target the  yield of $\tau$--events
for the Fermilab and CERN reference beams  in the 
high $|\Delta m^2|$  region  is well  described by:

\begin{equation}
\left ( {N_{\tau} \over N_p} \right)_{Fermilab} \simeq 16.5 ~ \sin^2 2 \theta ~
( {\rm kton} ~10^{19}\,{\rm pot})^{-1}
\end{equation}
\begin{equation}
\left ( {N_{\tau} \over N_p} \right)_{CERN} \simeq 126~ \sin^2 2 \theta ~
( {\rm kton} ~10^{19}\,{\rm pot})^{-1}
\end{equation}

Using  the beam intensity 
$N_p/10^{19} = 20$ (3)  for Fermilab  (CERN)
the    absolute rates of $\tau$--events  become:

\begin{equation}
N_{\tau}({\rm Fermilab}) \simeq 328  ~ \sin^2 2 \theta ~
( {\rm kton} ~{\rm yr})^{-1}
\end{equation}
\begin{equation}
N_{\tau} ({\rm CERN}) \simeq 383  ~ \sin^2 2 \theta ~
( {\rm kton} ~{\rm yr})^{-1}
\end{equation}
Note  how in this case the lower intensity of the CERN beam   
is overcompensated by the higher  proton energy.

In the low  $|\Delta m^2|$  region the  rate of $\tau$ production
for  the reference beams  for a  fixed  number of accelerated protons are:
\begin{equation}
\left ( {N_{\tau} \over N_p} \right)_{Fermilab} \simeq 0.11  ~ \sin^2 2 \theta ~
(\Delta m^2_{-3})^2  ~( {\rm kton} ~10^{19}\,{\rm pot})^{-1}
\end{equation}
\begin{equation}
\left ( {N_{\tau} \over N_p} \right)_{CERN} \simeq 0.28  ~ \sin^2 2 \theta ~
(\Delta m^2_{-3})^2  ~( {\rm kton} ~10^{19}\,{\rm pot})^{-1}
\end{equation}

\vspace {0.15 cm}
\noindent Using  the   predicted proton beam intensities
the    absolute rates of $\tau$--events  become:
\begin{equation}
N_{\tau}({\rm Fermilab}) \simeq 2.23  ~ \sin^2 2 \theta ~
(\Delta m^2_{-3})^2  ~( {\rm kton} ~{\rm yr})^{-1}
\end{equation}
\begin{equation}
N_{\tau} ({\rm CERN}) \simeq 0.84  ~ \sin^2 2 \theta ~
(\Delta m^2_{-3})^2  ~( {\rm kton} ~{\rm yr})^{-1}
\end{equation}
For $|\Delta m^2| \aprle  10^{-2}$~eV$^2$  the  oscillation  probability
$P \propto E_\nu^{-2}$   depresses the contribution of high energy neutrinos
and the softer  but more intense  Fermilab  beam  results in 
a $\tau$--rate that is approximately 2.6 times  larger than  the CERN beam. 
A  realistic  comparison between the two projects 
should  take into account the  
energy spectrum of the produced $\tau$ leptons, and take into account
the detection efficiency and sources of background.

\begin{figure} [t]
\centerline {\psfig{figure=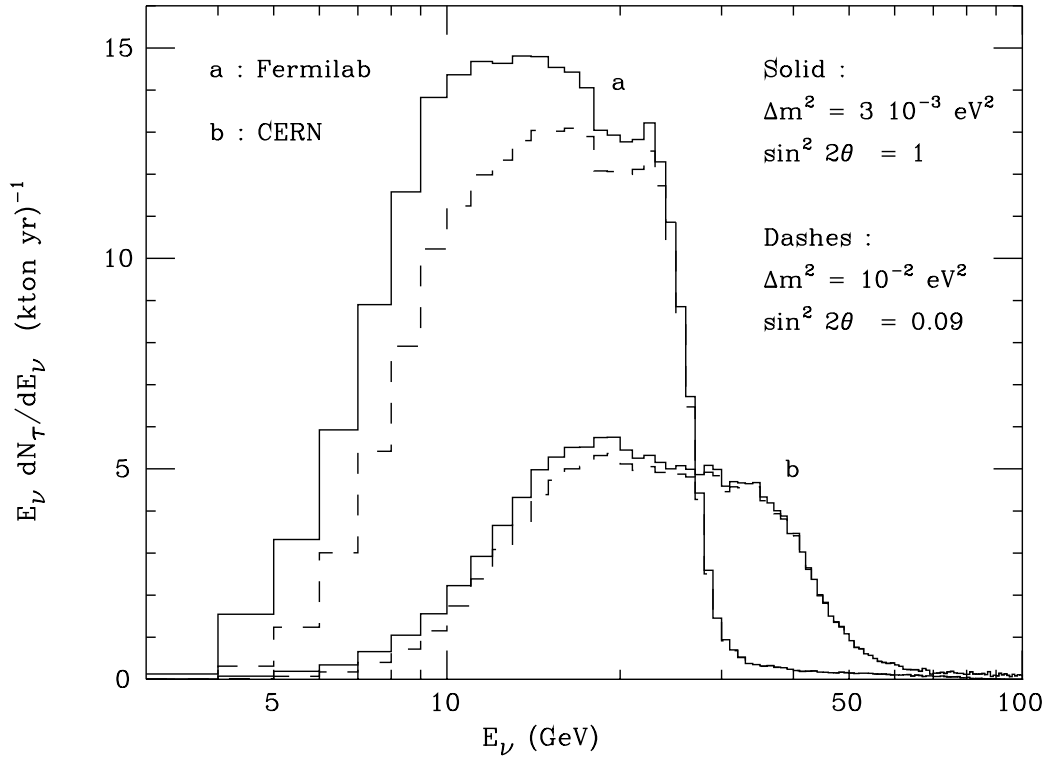,height=10cm}}
\vspace {0.4 cm}
\caption{ 
dN/dLog${E_\tau}$ 
distribution  of
the $\nu_\tau$'s interacting via  charged current
for the Fermilab to Minos  and the CERN to Gran Sasso
LBL neutrino  beams. The distributions  are calculated
assuming the existence of $\nu_\mu \leftrightarrow \nu_\tau$
oscillations 
with  $\Delta m^2 = 3 \times 10^{-3}$~eV$^2$  and  $\sin^2 2 \theta = 1$
(solid  histograms) and
$\Delta m^2 = 10^{-2}$~eV$^2$  and  $\sin^2 2 \theta = 0.09$.
The absolute scale  is in (kton~yr)$^{-1}$, assuming  a number of
protons of target of $2 \times 10^{20}$ and  $ 3 \times 10^{19}$
for  the two  LBL beams.
\label{fig:tau1} }
\end{figure}

\vspace {0.2 cm}
In fig.~\ref{fig:tau1}  we show the energy  distribution of the interacting 
$\nu_\tau$ neutrinos for  the Fermilab and CERN   LBL projects.
We have  made two hypothesis for the  $\nu$--oscillation parameters:
\begin {enumerate}
\item $\Delta m^2 = 3 \times 10^{-3}$~eV$^2$ and $\sin^2 2 \theta = 1$,
\item $\Delta m^2 = 10^{-2}$~eV$^2$ and $\sin^2 2 \theta = 0.09$.
\end{enumerate}
The  product  $(\Delta m^2)^2~\sin^2 2 \theta$
is equal    for both  choices of parameters.
This results in spectra of  $\tau$  leptons  that are
approximately equal in   shape and  normalization.
For large $E_\nu$ the spectra are identical, and only 
approaching the energy threshold 
the oscillation  probabilities  for the  two cases begin to  
differ appreciably.
The  integrated rates  for 
$\tau$ production rate  in units  of  (kton~yr)$^{-1}$ 
are 19.6 and 15.8 for the Fermilab beam  and  7.5  and 6.9 for
the CERN beam. Most of the difference is accumulated 
at low energies where the efficiency for identification is  smaller
and it will  be  very difficult  to discriminate between the
two  sets  of parameters  discussed in this example.
Note that both values of $|\Delta m^2|$ chosen in our 
examples are  larger  that the best fit  point
of Super--Kamiokande. For lower values of $|\Delta m^2|$  the {\em shape}
of the energy distribution of the $\tau$--events   becomes  constant.
In summary  the  energy spectrum of 
the $\nu_\tau$'s  at the far  detector
for  all values of $|\Delta m^2| \aprle 10^{-2}$~eV$^2$
(that is  in the region  indicated by Super--Kamiokande)
is  well described   with the asymptotic  form 
\begin {equation}
{dN_{\nu_\tau} \over dE_\nu } (E_\nu) = 
\sin^2 2 \theta \; 
(\Delta m^2)^2~
{L^2 \over 16} ~{ \phi_{\nu_\mu}^0 (E_\nu) \over E_\nu^2 }
\end{equation}
that has a shape  independent from  the oscillation parameters.
The measurement of the absolute value of the rate
of $\tau$  production does measure the product
$(\Delta m^2)^2 \, \sin^2 2 \theta$.
The  measurement of the shape of the 
energy  spectrum of the $\tau$'s  produced in 
the detector  can  confirm that indeed the  probability of 
flavor  conversion has
the  energy dependence predicted  by the theory, but this 
prediction is unique.  The only information on the parameters  is  obtained
from the absolute normalization.

Is the   predicted  rate  of $\tau$ events measurable ?
The  strategy for the detection of the signal  must find the right
balance between   mass and  resolution.  In general  
the number of  detected  $\tau$'s  can be written as:
\begin{equation}
N_\tau^{det} =  (\sin^2 2 \theta ~ |\Delta m^2|^2)
 ~M~t~N_p~{n_\tau^0 \over N_p}
~\langle \epsilon_{det} \rangle
\end{equation}
Where $M$ is the detector  mass, $t$ the running time,
$N_p$ the proton beam average intensity,  
$n_\tau^0$   (defined  in equation (\ref{eq:nutau0})) depends on the details
of the design of the beam line,
 and $\langle \epsilon_{det} \rangle$ is the  detection
efficiency  averaged  over the $\nu_\tau$ energy spectrum.

As  concrete   example of   an  experiment the capability
to identify   $\nu_\tau$ interactions  on an event by event  basis
we can  consider the OPERA  \cite{OPERA}  and ICARUS  
\cite{ICARUS} detectors\footnote{This  is not a comprehensive
review of the existing proposals. A more complete  discussion should
include  a discussion of  detectors  such as NOE and
MINOS that  have  larger mass   and  coarser
resolution.}.
The OPERA  detector  is based on a   sandwich  
(iron/emulsion/drift/emulsion) of  thin 
($\sim  1$~mm) iron plates  alternated
with   emulsion sheets  for high resolution  tracking  and  layers 
($\sim 2.5$~mm) of `drift space'   filled with  a very  low  density  material.
Events where a $\tau$  lepton is  produced  and  decays 
in the  first  drift space  after the vertex  can  be  identified
measuring the direction(s)  of the charged decay product(s) in the 
following  layers of  emulsions.
All $\tau$ decay modes  are observable, and the 
overall detection  efficiency taking (most of the inefficiency is 
due  to  $\tau$'s that decay inside  the iron)
is estimated as $\epsilon_{d} \simeq  0.35$.
A possible mass is $M \simeq 0.75$~kton.
OPERA  running at CERN  could  detect in 5 years a  number of 
$\tau$'s:
\begin{eqnarray}
N_\tau^{det} ({\rm OPERA} ) & =  & (\sin^2 2 \theta ~|\Delta m^2_{-3}|^2) ~
0.75 \times 5 \times 3 \times  0.28 \times 0.35 \nonumber \\
& =  & (\sin^2 2 \theta ~|\Delta m^2_{-3}|^2) ~1.1
\label {eq:OPERA}
\end{eqnarray}
with a background  smaller  than a single  event.
For the  best  fit  point of  Super--Kamiokande 
the  signal  predicted by equation
(\ref{eq:OPERA})   corresponds to 4.8  events, 
for   $|\Delta m^2|/10^{-3}$~eV$^2 = 0.5$ (6)
(the 90\% C.L. interval estimate  by SK) 
to   (0.28)  (40)  events.
In  the  first line  of equation (\ref{eq:OPERA})  we
 have written explicitly the   estimates of the
 most important  quantities  that   we have  used  for  the prediction.
A more intense proton beam,  
a  better  optimization of the focusing system,  
and  a higher detection efficiency are all  direction of improvement
not only possible but actively pursued.  
 Formula (\ref{eq:OPERA})  can of course be easily rescaled.  
Qualitatively,  we can  conclude  that  a 5--years long dedicated 
effort  based on this  emulsion--based  vertex detection technique  can  detect
a $\tau$ signal  (if $\nu_\mu \to \nu_\tau$ oscillations  are  indeed
the cause of the Super--Kamiokande data) in most  but    very likely 
not all the region of parameter space indicated 
by the  atmospheric  neutrino data.

The ICARUS  detector is  based on a new  technology  
that allows  to obtain  high  resolution (`bubble--chamber 
quality') images of the 3--D  deposition of ionization
in a large volume  of  liquid argon.  This detector 
will not have  the spatial  resolution to detect
the $\tau$  decay  vertex, however   the decay mode 
$\tau^- \to \nu_\tau \overline{\nu}_e e^-$  
(with a branching ratio  of 18\%) can  be detected
with  negligible  background  identifying  the electron
and the missing  transverse 
 momentum  due to the two  neutrinos in the  final state.
The overall efficiency  can  be roughly estimated  as  
$\langle \epsilon_{det} \rangle \sim 0.18\times 0.45 \simeq 0.08$.
The  lower  efficiency with respect to OPERA   will be  compensated by 
the larger  mass  for $M \aprge 3.2$~kton.

Higher  mass, lower  resolution  detectors  appear not very  likely to
have  a higher sensitivity  than  the examples we have discussed.

What is the conclusion of this discussion ?
If the  strategy  to develop a 
 high energy    long--baseline  neutrino beam and 
search for the appearance of  charged current  interactions of $\nu_\tau$'s
sufficiently promising to
deserve the  very high investment in  human and  financial   resources 
that are required ?  
We will  leave the  reader to make her/his own judgement.

\section {Disappearance Experiments}
The threshold for the charged current interaction 
of  $\nu_\tau$ on free  neutrons is  $E_t = 3.47$~GeV 
(a little  higher for bound nucleons).
At low  energy  the  cross section is  strongly suppressed by
kinematical effects and effectively  only $\nu_\tau$ 
with energy $E_{\nu} \aprge 4$--5~GeV can  effectively  interact  via  the 
charged current interactions.
For $|\Delta m^2|$ in the confidence interval of Super--Kamiokande 
and the path--length under  discussion,  the oscillation probability
of the neutrinos  well above threshold is small  and 
appearance experiments  are difficult.
The  oscillation probability   (for a fixed $L$)
reaches a maximum for a a discrete  set
of values of  the 
neutrino energy. For  the K2K  (Fermilab/CERN)  project
these values are $E_\nu = 0.20/n~\Delta m^2_{-3}$~GeV
($E_\nu = 0.59/n~\Delta m^2_{-3}$~GeV)   where  $n$ is  a positive integer.
For  maximal  mixing the flux of $\nu_\mu$'s at these  energies  will vanish.
In the SK 90\% confidence interval  $\Delta m^2_{-3} \le 6$
these energies are always   below the threshold
and it natural to explore the possibility to  measure the
oscillation probability  at low  energy.

The $\nu_\tau$  that  are below  threshold 
cannot  have a charged  current interaction, they   `disappear'.  
This `disappearance'  can be  detected  observing
a deformation of the  energy spectrum  of the CC interactions, 
or  measuring a  ratio N(nc)/N(cc) of    the numbers of
neutral  and  charged current  events  larger than  expectations.
This program  appears  straightforward but it is  not easy  to 
put in practice. We  can list  some of the problems of the 
experimental  strategy.

(1) It is difficult  to construct  a neutrino  flux  
that is  sufficiently intense at low  energy.  
The  neutrino flux  at the far detector
is proportional  (see section 2.2) 
to  a factor $(p_{\perp\nu}/E_\nu)^2$   because  the 
neutrinos are emitted in a cone that shrinks  with  increasing  energy.
An appropriate  optimization of the focusing system\cite{nice,Baldini}
can increase the low energy flux reducing the   angular  divergence of the
charged  mesons  of low energy\footnote{The low  energy mesons
are `over--focused' in the reference
designs  of the Fermilab and CERN designs.}
however the transverse momentum   due to the parent decay 
cannot  be avoided.
In a `standard' design  of the beam line the  approximation of the 
`perfect  focusing'   (see  fig.~\ref{fig:ideal}
and the discussions in section 2.2 and in the appendix)  
is the highest obtainable flux. It possible that  even 
this  maximum  flux have an intensity   below   what is necessary to explore 
the entire SK confidence interval.
Perhaps  a very innovative  design of the neutrino  beam line 
with a  `thick' target that  allows   the  reinteraction of high 
energy secondary particles  transforming the energy 
of the  primary protons in  a   large 
number of low  energy pions could  result in a  much  higher 
flux of low  energy neutrinos,  but this idea has not yet been  explored 
in any detail.

(2). The   method  of measuring  the ratio NC/CC of neutral current 
and charged  current events becomes  increasingly difficult 
for $E_\nu \aprle  1$~GeV  where  the lowest  multiplicity  channels
(elastic, quasi--elastic, single pion  production)  are    dominant
and the theoretical uncertainties   large.

(3) The energy spectrum of the interacting 
neutrinos in the low energy region will
be changing rapidly $\propto E_\nu^2$, and the deformations of the spectrum
not easy to measure.

(4) The systematic uncertainties about the no--oscillation flux,
the  background sources  and the detector response could 
become a limitation in the sensitivity of the technique.  
A near detector appears to be essential to reduce these systematic
uncertainties to the desired levels. 
However some  systematic  effects are likely to   remain
even  in  a  far/near  comparison.  The   two  detectors  cannot be
identical, and in  the near detector    only  events 
in   a sufficiently
small  region close to the beam axis  should be considered.

\vspace {0.3cm}
The sensitivity of the KEK  to Kamioka  project  \cite {K2K}
is   based on the  disappearance method, since essential all the neutrinos
produced by the 12~GeV proton beam will be below the 
threshold for $\tau$ production.  The experiment is 
expected to start taking data 
in january 1999, years  ahead of the Fermilab  and CERN beams.
The  neutrino mass interval 
that can be  excluded   at 90\% confidence level
after  3  years  of  running if no  signal is observed is
$|\Delta m^2| \aprle 3 \times 10^{-3}$~eV$^2$.

\section {Sensitivity of atmospheric neutrino experiments}

The  `atmospheric  neutrino  anomaly'
has  initially appeared as the  measurement of  a 
ratio  $R_{\mu/e}$  of  $\mu$--like and $e$--like events
lower  than the  MC expectation.
The scientific community   remained  very skeptical, after all
$R$  is  a single  number and  particle identification a  non--trivial task.
With  the measurement of  higher  energy
events  (the multi--GeV sample  of Kamiokande)\footnote{The 
reason  why the oscillation patterns
are  more easily  recognizable for higher energy events  
is  mostly due to the fact that
the  direction of the incident  neutrino is better
measured because the  detected charged lepton 
is  emitted in a cone  that   shrinks  with increasing  energy.}
and larger  statistical samples 
(22.5 kton of fiducial mass of  Super--Kamiokande)
it has been possible 
to  detect {\em patterns} in the energy and  zenith angle distributions
of the events that are precisely those  expected in the case
of neutrino  oscillations.

With the  detection of these patterns the hypothesis  that oscillations 
are  the explanation  of the  data   has become  much more  convincing.
It is in fact difficult to imagine what 
combination of systematic effects could  produce
the energy--dependent up--down asymmetry  observed by Super--Kamiokande
for  the $\mu$--like events.
In fact until now no alternative explanation (besides  neutrino
oscillations)  has  been  offered for these effects.

The  patterns  of the   distortions 
of the energy and   zenith angle  distributions  of the  detected  events 
have to be  estimated  comparing the data with  a theoretical prediction, and
one  has  to consider  carefully the uncertainty in the no--oscillation
predictions. However theoretical uncertainties  are not expected  to be the
dominant  source of systematic errors.
Two very  simple  and  very solid  predictions  for  the  neutrino
flux in the absence of oscillations: the  flavor ratio 
$\nu_\mu/\nu_e  \simeq 2$, and the 
up--down symmetry  (equations (\ref{osc-prob}) and
(\ref{osc-prob-asym})) 
   eliminate  most of the systematic  uncertainty about the prediction.

\vspace {0.3 cm}
In the case  of $\nu_\mu \leftrightarrow \nu_\tau$ oscillations
the oscillation probability  depends  only on  the ratio $L/E_\nu$,
and perhaps the most  physically intuitive and
transparent method  to    analyse  the data
is to study the distribution of events as a function of $L/E_\nu$.  
This method  has  been  discussed  in the past
and  recently vigorously advocated  by Picchi and Pietropaolo
in ref.\cite{Picchi-Pietropaolo1}. 

In the  presence of  oscillations the  distribution of events as a function
of $L/E_\nu$ is  distorted  by   a factor  that is simply
the oscillation probability  (eq.\ref{osc-prob}).   In any experiment
the finite   resolutions  in the measurement of the neutrino energy
and path--length smear  the measured  distribution $dN/d(L/E_\nu)$
and  partially  wash--out the   sinusoidal pattern 
of the oscillations.

\begin{figure} [t]
\begin{center}
\begin{tabular}{c}
{\psfig{figure=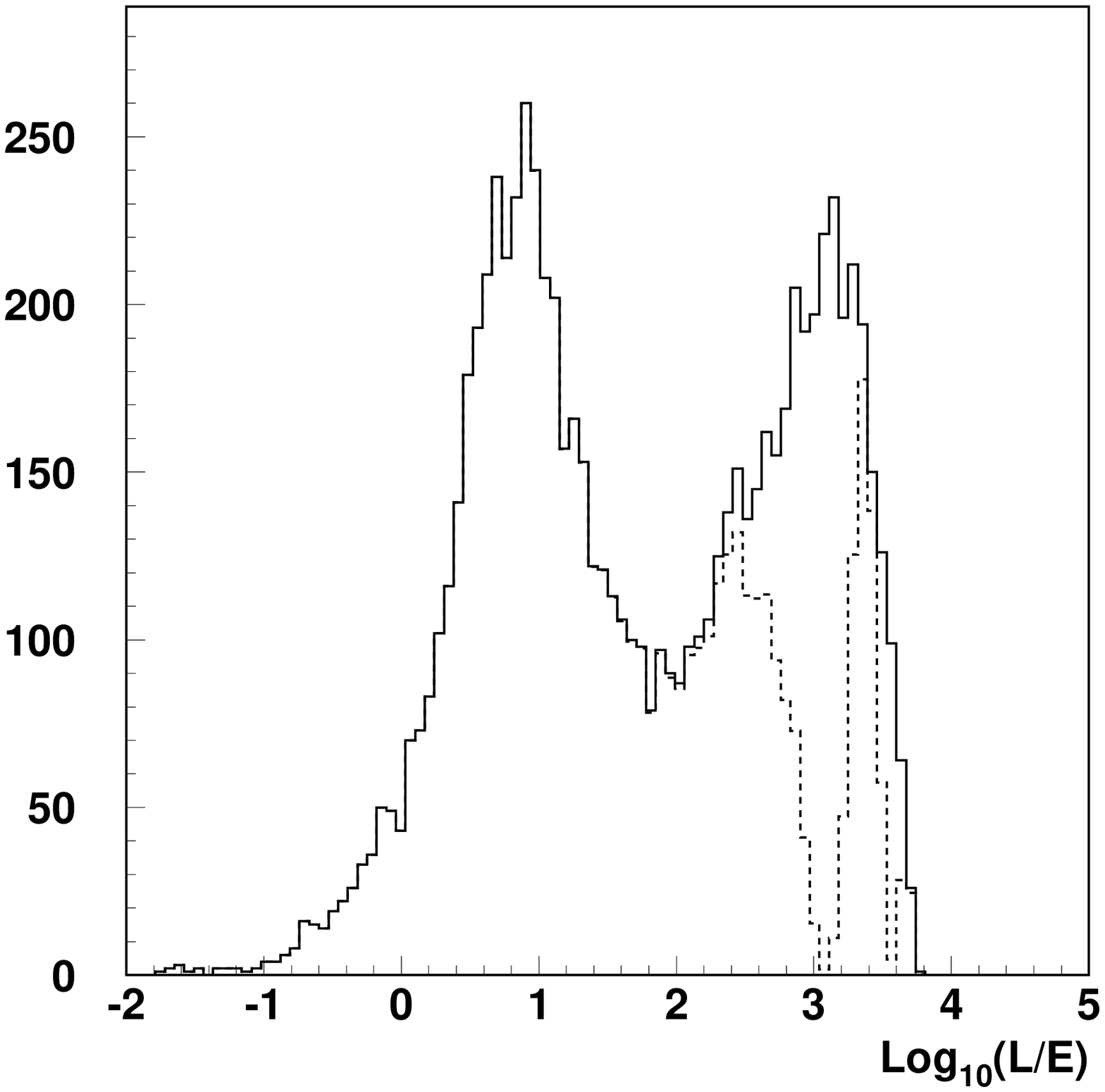,height=8cm}} \\
{\psfig{figure=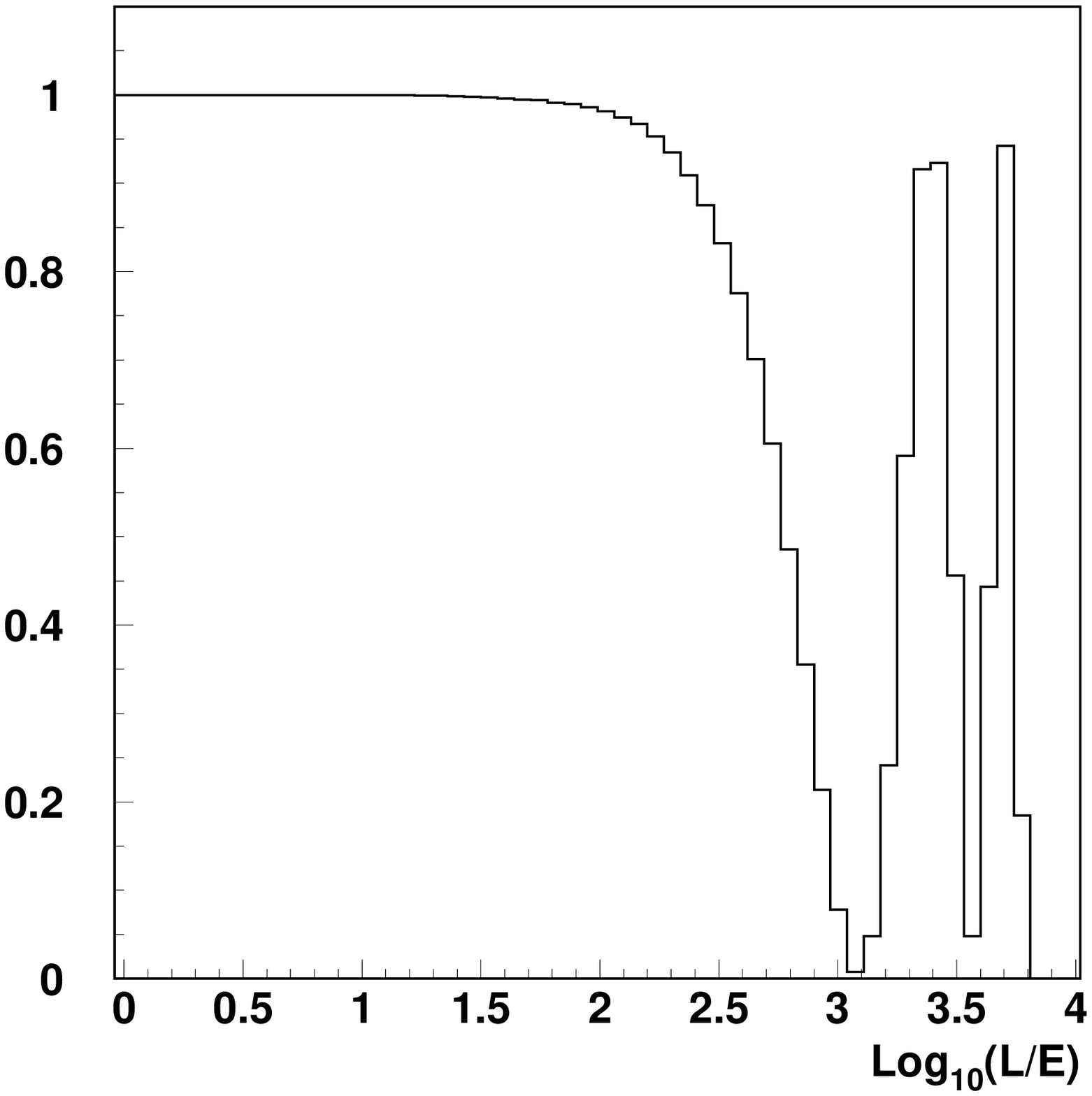,height=8cm}} 
\end{tabular}
\vspace {0.25 cm}
\caption{Upper panel: distribution in $L/E_\nu$, in units of km/GeV, of the
charged current  $\nu_\mu$ and $\nubar_\mu$ events   
expected  for  atmospheric neutrinos 
with a cut $p_\mu \ge 2.0$~GeV at the Kamioka site. The solid line corresponds
to no oscillation hypothesis, the dotted line is the expectation for
maximal mixing and $\Delta m^2$ = 10$^{-3}$. 
We are assuming  exact knowledge of
the direction and energy of the neutrinos;  only fluctuations in the
position of the  neutrino  creation point are  contributing
to the  resolution in $L/E_\nu$. 
Lower panel: ratio of the oscillation case to the no--oscillation hypothesis
as a function of $L/E_\nu$.
\label{fi:exampl1} }
\end{center}
\end{figure}

In the upper panel of fig.~\ref{fi:exampl1} we show a
montecarlo calculation of the distribution
of $L/E_\nu$  for the charged  current interactions  of
muon neutrinos and antineutrinos 
(with a cut $p_\mu \ge 0.5$~GeV) 
at the Kamioka  mine location.
The solid  line is the  prediction
in the absence of oscillations, the dashed line  is the  expectation
in the presence of $\nu_\mu \leftrightarrow \nu_\tau$ oscillations
with  maximal  mixing and 
$|\Delta m^2| = 10^{-3}$~eV$^2$.
The plot is calculated  for a `perfect  detector'
assuming that the   energy and direction of each  neutrino is reconstructed
with  infinite  precision.  Even in this case 
however the neutrino path--length $L$ is  not  perfectly measured
because the  neutrino creation point  is not known. 
For each   event the   oscillation probability is 
calculated using the neutrino energy 
and the `true'  $L$  generated  including  fluctuations \cite{hprod} in the
position of the $\nu$ creation point.  
To each event is  then assigned 
a `reconstructed' value of the path--length 
that is the most likely value  of $L$ for a given zenith angle.

In the lower panel  of  fig.~\ref{fi:exampl1}  we show the 
ratio of the oscillated plot to the no-oscillation hypothesis 
as a function of $L/E_\nu$  in a restricted interesting
range. 
The features of the oscillation phenomenon are  
unmistakable and  the measurement of its parameters straightforward.

\begin{figure} [t]
\begin{center}
\begin{tabular}{c}
{\psfig{figure=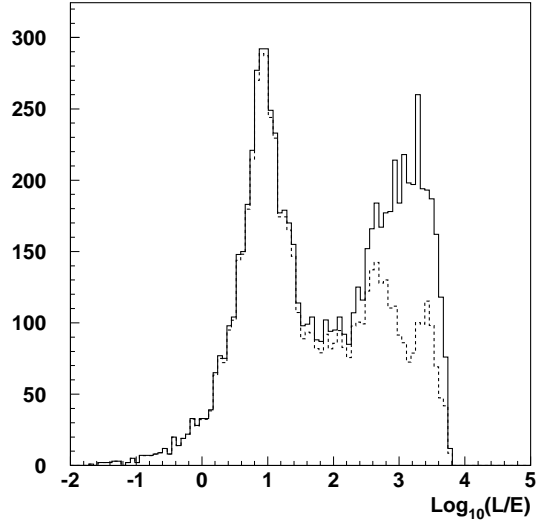,height=8cm}} \\
{\psfig{figure=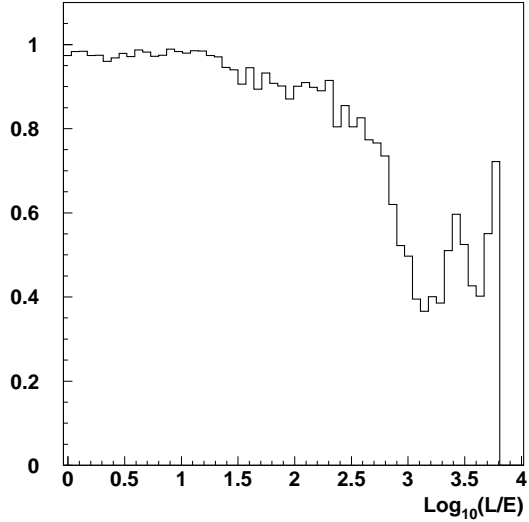,height=8cm}} 
\end{tabular}
\vspace {0.25 cm}
\caption{Upper panel:  distribution in $L/E_\nu$, in units of km/GeV, of the
charged current  $\nu_\mu$ and $\nubar_\mu$ events   
expected  for  atmospheric neutrinos 
with a cut $p_\mu \ge 2.0$~GeV at the Kamioka site. 
The solid line corresponds
to no oscillation hypothesis, the dotted line is the expectation for
maximal mixing and $\Delta m^2$ = 10$^{-3}$. 
The direction and energy of the  neutrino  are estimated
as $E_\nu \simeq E_\mu$ and $\Omega_\nu = \Omega_\mu$.
The  statistics of the montecarlo calculation corresponds  to  the large
exposure of 500~(kton~yr).
Lower panel: ratio of the oscillation case to the no--oscillation hypothesis
as a function of $L/E_\nu$.
\label{fi:exampl2} }
\end{center}
\end{figure}

Fig.~\ref{fi:exampl1}   represents of course an ideal case.
In any  realistic detector the   experimental resolutions will  
partially wash--out the spectacular  pattern.
A more realistic  example, that 
is a rough approximation  of Super--Kamiokande  is  shown in
fig.~\ref{fi:exampl2}.  We have 
selected  `single  ring events  contained'  events 
(always with the  cut $p_\mu \ge 2$~GeV), using the Cherenkov threshold  for  
charged  particle  detection, and  the geometry of  the detector
and the muon range in water for the containment, and 
estimated the energy and  direction of the neutrino with 
$\Omega_\nu = \Omega_\mu$ and  $E_\nu = E_\mu$
(we have neglected  the  error  measurement in
$E_\mu$ and $\Omega_\mu$, since  the  resolutions  are  smaller
than the intrinsic fluctuations due  to the   production cross section
this is  a reasonable approximation).
One can  see that in this case the oscillation patterns  are  
less spectacular than  in the ideal case but nonetheless quite clear.

Note that after the inclusion of the detector
resolutions the  shape of the suppression factor  due to oscillations
depends on the minimum muon energy considered $E_\mu^{\rm min}$.
With increasing $E_\mu^{\rm min}$ the  
pattern of damped oscillations becomes more  clear, because 
the  muon direction is more strictly correlated with the neutrino direction
and $L$ is better  determined,
however the  number of  detected events decreases  because of the
steepnes of the atmospheric neutrino  energy spectrum.
In a complete analysis  all  events  shouls  be included, taking into account
the energy dependent experimental  resolutions.

Fig.~\ref{fi:exampl2}  has  been  calculated   considering a very large
exposure  (approximately 20 years of Super--Kamiokande) in  order
to show   clearly the oscillation patterns. For  a lower   realistic exposure 
the statistical fluctuations  will be larger  and the   determination of the
oscillation parameters less precise.
The claim of `evidence'  for  neutrino oscillations  in atmospheric  
neutrino data  by the Super--Kamiokande collaboration is 
is based on 1.5~years. Additional data should allow  a more clear  detection
of the oscillation patterns  and a better determination 
of the parameters.

Looking at fig.~\ref{fi:exampl2}, 
one can note immediately see  how 
the oscillation parameters  $\Delta m^2$  and $\sin^2 2 \theta$ are
determined.
The suppression factor  due   to oscillation
is unity  for small $L/E_\nu$,   decreases
with increasing $L/E_\nu$ reaching a minimum
at $L/E_\nu \sim 2\pi/|\Delta m^2|$ and  then 
shows  some `damped   oscillations' 
around the  average value $\simeq 1 - {1\over 2} \sin^2 2 \theta =  0.5$. 
These  features    will be  true  in general.
The path--length  of  atmospheric  neutrinos  is  in the range
$L \simeq 10$--$10^4$~km, and  assuming  an  energy   interval
$E_\nu \sim 0.1$--10~GeV  the  range  of
$L/E_\nu$  is $\sim 1$--10$^5$~km/GeV.  Given  this very  large   range
of $L/E_\nu$,  it is  possible 
(at least in the region $|\Delta m^2| \simeq 10^{-4}$--$10^{-1}$~eV$^2$)
to distinguish  three regions  of 
$L/E_\nu$  depending on ($|\Delta m^2| \;L/E_\nu$) being much smaller,
larger or comparable   with unity.
\begin {enumerate}
\item For small  $L/E_\nu$
the oscillation   cannot  develop.
\item  For large $L/E_\nu$ range the oscillations  are rapid
and  difficult to see for  realistic  detector resolutions. The  average 
value of the  suppression factor is $1 - {1\over 2} \sin^2 2 \theta$.

\item  In the intermediate  region
the  oscillating  pattern is more evident. This region  is  the crucial one 
for the determination of $\Delta m^2$.
Of particular  importance is 
the  shape of  the distribution around  the value  
$L/E_\nu \sim 2\pi/\Delta m^2$   (corresponding to
the point where the  oscillation  probability 
has the first maximum).  The  identification of a maximum in the suppression 
factor would be at the same  time a   clear  proof of the existence 
of oscillations  and  a   determination of the  $|\Delta m^2|$ value.
\end{enumerate}

\subsection {A new detector for atmospheric neutrinos ?}
Most of the  data on atmospheric neutrinos has  been  obtained 
with water \v{C}erenkov detectors. 
Recently an  iron  calorimeter, the Soudan-II
detector has obtained a result that 
supports  the SK result.  However  this  detector 
has a fiducial  mass of only 0.5~kton   and will not be able  to collect
a very large statistics.
The clear confirmation of the atmospheric neutrino  anomaly
using a different  experimental technique  is  clearly  very  desirable.

Is it possible  to  develop a  detector  for  atmospheric neutrinos
with the capability  not only to confirm the Super--Kamiokande result
but to  provide   a measurement that is
at least in some respects  of superior  quality ?
This  task is  not easy,  because of the remarkable  qualities 
of the Super--Kamiokande  experiment: very large mass,
very good spatial and energy resolutions, isotropic  efficiency.

A possible direction   is the  development of a moderate mass
but higher  resolution detectors   capable  
for example the to   measure  also the  nuclear recoil
in quasi--elastic  neutrino  scattering.  The ICARUS  \cite{ICARUS} 
detector could have 
the  potential to perform this  measurement, however   reinteractions 
in the target nucleus  could  become  the    most important source
of error  in the determination of the   energy and direction of
the incident neutrino \cite{nuclear-effects}.

\vspace {0.4 cm}
Recently \cite{Picchi-Pietropaolo1,Picchi-Pietropaolo2,nice}
the potential of a large mass  tracking calorimeter  for the measurement 
of atmospheric  neutrinos has also been  discussed  as an attractive 
possibility.
A possible  advantage of  a calorimeter compared  to
a water Cherenkov detector is the potential  to perform 
a better  measurement   of the
events with several particles in the  final state
because of superior pattern recognition  capabilities.
The possibility  to  include a  magnetic  spectrometer \cite{nice}
for the measurement of the momentum of the muons
that  exit the detector  in the `semi--contained' events  
has   also been  investigated.
The  optimum design  for such a  calorimeter (with the right  balance between 
mass and granularity) is still under  discussion.

A natural  idea  is of course to try to 
do two things at the same time
and develop  a detector that  can  at the same time perform a  high quality
measurement  of atmospheric  neutrinos  and of the neutrinos of a
LBL  beam. The idea is  attractive, 
but the design of  such a detector is not  easy, because 
the optimization of the detector performance for the two  type  of  measurements
can push toward different  non--compatible solutions.
As an obvious  example, the neutrinos  
from the beam  come from  a single (horizontal) direction
while  atmospheric  neutrinos  are  quasi--isotropic, with  the vertical  axis 
as the  most  important direction. The orientation of the elements of the 
detector is  therefore  problematic.

\section {Discussion and conclusions}

The  discovery of neutrino oscillations  could open
a precious   new window  on the physics beyond the standard model.
It is therefore necessary to have an independent and  unambiguous 
confirmation of  the evidence  collected by Super--Kamiokande
 and supported by  other  experimental  results. 
The next step is to measure with precision
the oscillation parameters.

What is the  best strategy  to achieve  these  goals ?
Long-baseline   neutrino beams  have  been proposed  since  several years
as   a sensitive instrument to  perform  detailed   studies 
on neutrino oscillations.  However   the 90\% confidence  interval
obtained  by Super--Kamiokande 
($ 5 \times 10^{-4} < |\Delta m^2|  < 6 \times 10^{-3}$~eV$^2$)
is  lower than  earlier estimates and   as a consequence
the long baseline experimental programs have   become much more 
difficult and  less attractive. 
Do  they remain  the wisest  experimental strategy   
for a confirmation and  for
detailed  studies  of the phenomena  indicated  by  SK ?
The answer is not easy, but this  question should be studied 
in depth because of the intrinsic scientific interest of the problem,
and because in any case the new experimental  studies will 
require a very large investment of scarce human and financial resources.

In this  work we have  briefly discussed three different  
experimental approaches   to the problem:

1. a LBL `appearance'  experiment,

2. a  LBL `disappearance'  experiment,

3. a new  atmospheric  neutrino experiment.

\vspace {0.3 cm}
In a $\tau$--appearance experiment    the signal is  produced 
by  neutrinos  
above the threshold  for $\tau$ production, that is 
with $E_\nu \aprge 4$--5~GeV.
 For $|\Delta m^2|$ in the SK   confidence interval
the probability of oscillations 
of these  neutrinos is always small  and the
detectable  signal   difficult   to observe,  For the `reference'
design  of the Fermilab (CERN) beam, 
assuming maximal mixing  and  
$|\Delta m^2| = 10^{-3}$~eV$^2$,  the 
inclusive rate   of $\tau$ production  (assuming $\epsilon_{det}=1$) is
2.2~(0.84)~events/(kton~yr).
To a very good approximation  the signal is 
proportional  to $\sin^2 2 \theta ~(\Delta m^2)^2$. 
Considering the predicted event rate, and realistic  estimates 
of detector mass  and efficiency   one can conclude 
that a positive  signal can be detected   by
very sensitive detectors   but only in
part of the SK  confidence interval.  Improvements in beam intensity
and design, a larger detector  mass,   higher  efficiency   and  of course
more  patience in collecting data  can push 
the sensitivity   down to lower values  of $|\Delta m^2|$. However 
even  in the presence of  a negligible  background,  since 
the signal  is $\propto (\Delta m^2)^2$  the increase in sensitivity can only
come at a high cost.

\vspace {0.3 cm}
The feasibility of a disappearance  experiment 
capable of covering the entire Super--Kamiokande  confidence interval
has not yet been demonstrated and requires  additional studies.
The main problems  are: (i)  the  design of 
a neutrino  beam with the required intensity in the low energy region;
(ii) the demonstration   that  the  systematic  uncertainties 
in the  knowledge  of the no--oscillation beam  and in the detector 
response  can be  kept  below  the required level.
The KEK to Kamioka  (K2K) project   is  expected to start taking data 
in january 1999.  In the absence of a positive signal,
after  3  years  of  running,
it  would  exclude    at the 90\% confidence level
the  neutrino mass region $|\Delta m^2| \aprle 3 \times 10^{-3}$~eV$^2$.
The  results from K2K can be a very important guide for  the
other experiments.

\vspace {0.3 cm}
If the effects  observed in the measurements of atmospheric  neutrinos
are indeed  due to neutrino oscillations,
Super--Kamiokande, after  the analysis of  a larger 
sample of data, should be able to  see more clearly
the oscillation patterns, narrowing the 
confidence interval for the oscillation parameters.
The  confirmation of these results  by an independent  experiment
of  comparable sensitivity and  using a different  technique   
is very desirable.
The  design and construction of a neutrino  detector with 
the capability of performing measurements  of atmospheric  neutrinos
of comparable or superior  quality with respect to
Super--Kamiokande is not an easy task. 
The concept  of a large mass, high resolution tracking calorimeter 
is  under  investigation.

\section*{Acknowledgements}

Part of this work was presented by one of us (P.L.) during the
Workshop on ``Frontier objects in Astrophysics and
Particle Physics'' held in Vulcano (Italy) 
in may 1998.  This expanded written  version was  
conceived during  after session  discussions.
We wish to thank the organizers of the conference 
for having provided the stimulating atmosphere.
We acknowledge many illuminating discussions with colleagues,
that occurred also before and after the conference.
In particular we would like to thank 
U.Dore, 
A.Ereditato,
A.Ferrari, 
G.Giannini, 
S.Mikheyev,
G.Mannocchi, 
P.Picchi, 
F.Pietropaolo,
S.Ragazzi, 
A.Rubbia,
R.Santacesaria,
E.Scapparone, 
T.Stanev,
M.Spinetti, 
T.Tabarelli.

\newpage
\section*{Appendix: }
\section*{Analytic calculation of a  LBL $\nu$ beam}

The detailed  calculation of  a LBL neutrino  beam   is  a task that can 
only be performed with detailed  montecarlo codes,  however to
obtain {\em understanding}   about the possible characteristics  of
 a long baseline    neutrino  beam, it  is simple and useful  to  compute 
its  spectrum analytically.
We  can start   observing that  at the detector   site the
neutrino  creations  region is well  approximated as a point,
and the solid  angle    subtended  by the detector  
$\Delta \Omega_{det} = A_{det}/L^2 \simeq  10^{-10}$~radiants
($L$ is the neutrino path--length)  is small  and therefore
the neutrinos  to a good approximation  can  be  considered as
collinear, and  we can define a neutrino flux  at the detector
\begin{equation}
\phi_\nu (E_\nu) \simeq {\Delta \Omega_{det} \over A_{det}} ~
\left [ {d N_\nu \over dE_\nu \, d\Omega_\nu}  (E_\nu, \Omega_\nu)
\right]_{\Omega_\nu = \Omega_{det}}
\end{equation}
where $\Omega_\nu$ is the neutrino  direction and 
$\Omega_{det}$  is the direction of the   line of sight from
the accelerator to the detector. 
For a  perfect  beam design $\Omega_{det} = \Omega_p \equiv 0$
($\Omega_p$ is  the proton beam direction).

The   dominant source if $\nu_\mu$ 
(or $\overline{\nu}_\mu$ neutrinos 
is the decay  of charged pions and kaons.
In general we can write :
\begin{eqnarray}
{d N_\nu \over dE_\nu \, d\Omega_\nu}  (E_\nu, \Omega_\nu) & = &
\sum_{a =\pi,K} B_{a \to \mu\nu} \,\int dE_a \, \int d\Omega_a
\,{ dn_a \over dE_a d\Omega_a} (E_a, \Omega_a, ~E_0)
~P_{dec}(E_a)~ \nonumber \\
& &  ~~~\int d\Omega_{a\nu} ~{ dn_{a\to \nu} \over dE_\nu d\Omega_{a\nu}}
 (E_\nu, \Omega_{a\nu},~E_a)
~\delta [\Omega_{\nu} - (\Omega_a \oplus \Omega_{a\nu}) ]
\label{eq:beam-general}
\end{eqnarray}
where 
$dn_a/(dE_a d\Omega_a)$ is the  differential  distribution
in energy and  direction of the parent  mesons,
$P_{dec}(E_a)$ is the decay probability, 
$dn_{a\to\nu}/(dE_\nu d\Omega_{a\nu})$
is the distribution in  energy and  angle (with respect to 
the parent  direction) of the neutrino  produced  by  a primary meson
of  energy $E_a$, and  the  delta  function  imposes that
the  produced  neutrino has the  desired  direction.
The  neutrino distributions  with respect to a parent 
of given  energy and direction  are 
easily obtained, knowing that the 
$a\to \mu\nu$ decay  is isotropic  in the parent meson rest frame.
Using the conservation  of 4--momentum:
\begin {equation}
~{ dn_{a \to \nu} \over dE_\nu d\Omega_{a\nu}} (E_\nu, \Omega_{a\nu}, ~E_a)
= {1 \over 4 \pi} \; \int_{-1}^{+1} \;dz~
\delta [E_\nu  - \gamma \,p^*_a (1 + \beta\, z) ]
~\delta \left [\cos \theta_{a\nu} - {z + \beta \over 1 +  \beta \,z} \right ]
\end{equation}
where $z = \cos \theta^*$ is the cosine of the   c.m. decay  angle.
$p^*_a$ is the 3--momentum of the decay  products in the 
parent rest  frame  and $\gamma$ and $\beta$ are the Lorentz  factor and
velocity  of the parent  meson.

The simplest case is  the so called `perfect focus'  approximation,
that is the  assumption that 
all particles  with charge of the correct sign
after the  focusing system are  collinear  with the beam.
In this  case all  integrations in equation (\ref{eq:beam-general})
 can  be solved using the Dirac  deltas:
The solution   for the  most interesting case of   a detector  aligned
with the  proton beam 
(using  for  simplicity  the expression   valid for
ultrarelativistic  mesons)  is :
\begin{equation}
\phi_\nu (E_\nu; E_0, L) = 
{1 \over 4 \pi \, L^2} ~\sum_{a =\pi,K} {B_{a \to \mu\nu}  \over 1 - r_a}
~\left [ {dn_a \over dE_a}  ( {E_a},E_0 ) 
\; \exp \left ( - {T \,m_a \over \tau_a\, E_a} \right ) 
\right ]_{E_a = E_\nu/(1-r_a)}
~\left ( {E_\nu \over p^*_a} \right)^2
\label{eq:perfect-focus0}
\end{equation}
where $r_a = 1-(m_\mu/m_a)^2$  ($r_\pi = 0.5731$, and $r_K = 0.0458$),
and  $T$ is  the length of the tunnel.
Equation (\ref{eq:perfect-focus0}) has   a  transparent  meaning.
A  neutrino   of energy $E_\nu$     along the beam  direction
can  be produced  only  by a parent pion (kaon)  of  energy
$E_\nu = E_\pi/(1-r_\pi)$   that  decays   emitting 
the neutrino  is the forward direction.  The flux  of  neutrinos
of  energy  $E_\nu$ is  therefore  given  by the number of parent  mesons
of appropriate  energy
produced  in an  interaction,  multiplied  by the decay probability,
and  by a Jacobian  factor $(1-r_a)^{-1}\,(E_\nu/p^*_a)^2$, 
where $p_a^*$ is the
momentum of the neutrino  in the rest frame of the parent  particle
($p_\pi^* = 29.8$~MeV, $p_K^* = 236$~MeV). The Jacobian  factor
takes  into account the facts that    the range of neutrino  energy
is compressed  by  a factor
$1-r_a$ with respect to  the range  of  parent mesons,
and that the neutrinos  
are  produced  in a   solid  angle  that  shrinks with  increasing energy
as $E_\nu^{-2}$,  correspondingly the neutrino flux  at the
detector is  enhanced by the inverse factor.
Comparing the contribution of  pions  and kaons to the neutrino flux
we can notice that the kaon contribution is enhanced  because
of the  shorter lifetime, but  depressed because of the 
larger $p^*$.
In  the decays of  pions     (kaons) the  neutrino  can  take  at
most  a fraction  $1-r_\pi = 0.427$   ($1-r_K = 0.954$)  of the  parent energy,
therefore  all neutrinos   above the energy $0.427\;E_0$ are produced
by kaon decay. 

For  a detector  aligned  with the proton beam  but for  an arbitrary angular 
distribution of the final  state  particles
the neutrino  flux  can be  obtained  with a single integration,
for example as: 
\begin{eqnarray}
\phi_\nu (E_\nu; E_0, L) & =  &
{1 \over 4 \pi \, L^2} ~\sum_{a =\pi,K} \; {B_{a \to \mu\nu}  \over 1 - r_a}
\int d\cos\theta_a
~\left [ {d^2n_a \over d\cos\theta_a \,dE_a}  ( E_a, \cos \theta_a;~E_0)
\right. \nonumber \\
 & & ~~~~~~~ P_{dec} (E_a) ~\left ( {2 E_a \over p^*_a} \right)^2 
\left. 
~\left ( {E_\nu \over E_a (1-r_a)} \right) \right ]_{E_a = E_a(E_\nu, \theta_a)}
\end{eqnarray}
where 
\begin {equation}
E_a (E_\nu, \theta) =  {E_\nu \over 1- r_a} \left ( 2 \over 1 + \sqrt{ 1 -
(\theta \,E_\nu/p^*_a)^2} \right )
\end{equation}
is the parent    meson  energy that can  produce  a  final  state  neutrino
of energy $E_\nu$  at an angle $\theta$  with respect to the parent direction.

As far as the calculation of the number of  pions and kaons of energy $E_\pi$ 
($E_K$)
produced in an interaction  by a primary proton of  energy $E_0$ is concerned
(see eq.~\ref{eq:scaling}), 
in our analytic  calculation,
we have used  the form
$F(x) = C\,e^{-a x}\, (1-x)^b$  to fit  montecarlo generated data.

\newpage


\end{document}